\documentclass[fleqn,usenatbib]{mnras}

\usepackage[T1]{fontenc}

\DeclareRobustCommand{\VAN}[3]{#2}
\let\VANthebibliography\thebibliography
\def\thebibliography{\DeclareRobustCommand{\VAN}[3]{##3}\VANthebibliography}

\usepackage{graphicx}	
\usepackage{amsmath}
\usepackage[version=4]{mhchem}
\usepackage{multirow}
\usepackage{newtxtext,newtxmath}
\usepackage{natbib}
\usepackage{array} 
\usepackage{booktabs}
\usepackage{comment} 
\usepackage{endnotes}
\usepackage{xspace}

\newcommand{\kms}        {\ifmmode{\rm \,km\,s^{-1}}\else\,km\,s$^{-1}$\xspace\fi}
\newcommand{\unitNHI}    {\ifmmode{\rm \,cm^{-2}}\else\,cm$^{-2}$\xspace\fi}

\newcommand{\Lya}       {\ensuremath{\rm Ly\alpha\,}}
\newcommand{\CIV}       {\ensuremath{\rm C\,IV\,}}
\newcommand{\HeII}       {\ensuremath{\rm He\,II\,}}
\newcommand{\OIII}       {\ensuremath{\rm O\,III] \,}}
\newcommand{\NCIV}    {\ifmmode{N_{\rm CIV}}\else $N_{\rm CIV}$\xspace\fi}  
\newcommand{\RCIV}    {\ifmmode{R_{\rm CIV}}\else $R_{\rm CIV}$\xspace\fi}  
\newcommand{\sigmaR}       {\ensuremath{\sigma_{\rm R} \,}}
\newcommand{\vout}       {\ensuremath{V_{\rm out} \,}}
\newcommand{\sigmaint}       {\ensuremath{\sigma_{\rm int} \,}}
\newcommand{\tauo}       {\ensuremath{\tau_{\rm 0} \,}}

\newcommand{\vsep}       {\ensuremath{\Delta V_{\rm K,H} \,}}
\DeclareMathOperator\erf{erf}

\title[Tracing warm gas with C IV]{Tracing warm gas through C IV radiative transfer}

\author[Lim et al.]{
Jin Lim,$^{1,2}$
Seok-Jun Chang$^{3}$
\thanks{Corresponding Author, E-mail: sjchang@mpa-garching.mpg.de},
Max Gronke,$^{4}$
Fabrizio Arrigoni Battaia,$^{3}$
Hee-Won Lee,$^{5}$
and Kwang-Il Seon$^{1,2}$
\\
$^{1}$Korea Astronomy $\&$ Space Science Institute, 776 Daedeokdae-ro, Yuseong-gu, Daejeon 34055, Republic of Korea\\
$^{2}$Astronomy and Space Science Major, University of Science and Technology, 217, Gajeong-ro, Yuseong-gu, Daejeon 34113, Republic of Korea\\
$^{3}$Max-Planck-Institut für Astrophysik, Karl-Schwarzschild-Str. 1, D-85748 Garching, Germany\\
$^{4}$Astronomisches Rechen-Institut, Zentrum für Astronomie, Universität Heidelberg, Mönchhofstraße 12-14, D-69120 Heidelberg, Germany \\
$^{5}$Sejong University, 209 Neungdong-ro, Gwangjin-gu, Seoul, 05006, Republic of Korea
}

\date{Accepted XXX. Received YYY; in original form ZZZ}

\pubyear{\the\year{}}

\begin{document}
\label{firstpage}
\pagerange{\pageref{firstpage}--\pageref{lastpage}}
\maketitle

\begin{abstract}
The C~IV $\lambda\lambda1548,1551$ resonance doublet is a key tracer of warm gas ($T\sim10^5\,{\rm K}$) within and around galaxies. Recent observations have detected this line in both absorption and emission, revealing asymmetric profiles in galaxies and spatially extended haloes around active galactic nuclei (AGNs). Resonance scattering can strongly modify the emergent spectra and spatial distributions, complicating their interpretation. Using 3D Monte Carlo radiative transfer simulations, we study C~IV resonance scattering over a broad range of column densities, intrinsic emission-line widths, and outflow velocities. We find that multiple scattering broadens the line profile and, in outflowing media, modifies the doublet ratio, $R_{\rm CIV}$, defined as the flux ratio of the K and H components at 1548 and 1551\,\AA, respectively. When the outflow velocity approaches or exceeds the doublet separation ($\simeq500\,{\rm km\,s^{-1}}$), K-line photons are redistributed around the H component, driving $R_{\rm CIV}$ below its intrinsic value and, in optically thick fast outflows, even below unity. We also combine photoionization models with resonance scattering to investigate extended C~IV haloes around AGNs and compare them with He~II $\lambda1640$ emission. Simple photoionization models do not produce C~IV emission more extended than He~II, whereas resonance scattering redistributes locally produced and central-source C~IV photons to larger radii. These results demonstrate that the C~IV doublet ratio and spatial distribution provide complementary diagnostics of warm gas.
\end{abstract}

\begin{keywords}
radiative transfer -- scattering -- line: profiles -- galaxies: active -- galaxies: haloes 
\end{keywords}

\section{Introduction}

Galactic winds driven by star formation and active galactic nuclei (AGNs) are ubiquitously observed across cosmic time and exhibit multiphase structure extending over several kpc \citep[e.g.,][]{Veilleux2005}. These outflows play a crucial role in galaxy evolution by mediating interactions between the host galaxy and its circumgalactic medium (CGM), the vast gas reservoir within the host-galaxy halo.
They eject fuel for star formation into the CGM or intergalactic medium, regulate the accretion of gas from the CGM into the interstellar medium, and enrich the accreted material \citep{Faucher2023,Thopson_Heckman2024}.

Resonance doublet lines are powerful diagnostics of galactic winds and the CGM. The different ionization energies of these lines have been used to investigate the multiphase structure of gas in galactic outflows and the CGM \citep[e.g.,][]{Steidel2010, Heckman2017}. 
\Lya and low-ionization atoms such as Mg~II ($\sim 15.0\,$eV) trace cold gas ($T \sim 10^4\,$K), and high-ionization atoms such as \CIV, N~V, and O~VI ($\geq 47.9\,$eV) trace warm and hot gas ($T \geq 10^5\,$K).
In particular, the \CIV resonance doublet at $\lambda\lambda 1548, 1551$ is a key tracer of warm gas at $T \approx 10^5\,$K, where \CIV fraction peaks. As a result, \CIV serves as an important coolant and tracer of gas at $T \sim 10^5\,$K \citep[e.g.,][]{Tumlinson2017}.

The \CIV resonance doublet has been studied both in absorption and emission. From absorption line studies, \CIV has been reported with typical column densities of $N_{\rm{CIV}} = 10^{13 - 15} \unitNHI$ \citep[e.g.,][]{codoreanu2018,Rudie2019,Davies2023}, in reasonable agreement with cosmological simulations \citep{Garza2024}, and it has been observed out to impact parameters of $\sim100\,$kpc \citep[e.g.,][]{bordoloi2014,Liang2014,Dutta2021}.
Strong \CIV emission has been detected in sources ranging from stellar objects \citep[e.g.,][]{Feibelman1983,Michalitsianos1988,Michalitsianos1992,Than2024} to nebulae around star-forming galaxies and AGNs \citep[e.g.,][]{Heckman1991,Fab2015a,Fab2015b,Stark2015,Berg2019A,Travascio2020,Guo2020,Fossati2021,Matthee2022,Topping2024}. Through the detection of \CIV emission, the spatial distribution and kinematics of diffuse warm gas can be investigated without the need for bright background sources.

The detection of \CIV emission raises an interesting question about the doublet ratio of \CIV$\lambda 1548$ and $\lambda 1551$ ($F_{1548}/F_{1551}$). In particular, highly ionized star-forming galaxies exhibit strong nebular \CIV emission \citep[e.g.,][]{Berg2019B,Izotov2024}, with the two doublet components clearly resolved, making them ideal targets for measuring the doublet ratio.
Theoretically, the intrinsic doublet ratio is 2 due to their statistical weights of the transitions. 
However, the observed \CIV spectra reveal variations in doublet ratios, with a wide range of values from 0.3 to 2.4 \citep[e.g.,][]{Stark2015,Vanzella2016,senchyna2017,Tang2024}.
For instance, \citet{Topping2025} reported a ratio below unity ($F_{1548} / F_{1551} = 0.8$), where the spectrum is characterized by a strong \CIV $\lambda1551$ and a weak $\lambda1548$. This is different from the intrinsic profile of the \CIV doublet emission. The authors presumed that radiative transfer processes may modify the intrinsic doublet ratio during photon escape.

Recently, \CIV emission has also been detected as extended nebulae out to several tens of kpc around AGNs with extended \HeII $\lambda1640$ nebulae \citep[e.g.,][]{Guo2020,Fossati2021,Sabhlok2024,Galbiati2026}. 
Given the similar high-ionization energies of \CIV and \HeII (47.9 and 54.4\,eV, respectively), these lines are often detected together in the warm CGM at $T \approx 10^5$\,K.
Observations also show that \CIV emission is usually brighter and more spatially extended than \HeII at current surface-brightness limits. 
For example, \citet{Galbiati2026} detected extended \CIV and \HeII nebulae around a $z=3.66$ quasar with sizes of $\approx55$ and $\approx45$\,kpc and \CIV/\HeII ratios well above unity. 
In addition, \citet{Travascio2020} reported extended \CIV emission out to $\sim75$\,kpc in the absence of \HeII emission.
To generate these high-ionization emission lines, a hard radiation field is required so that photoionization by a powerful central source has traditionally been considered the primary mechanism. 
However, several studies have shown that pure photoionization models fail to reproduce the observations unless crude approximations of resonance scattering are taken into account \citep{Fossati2021,Obreja2024}.
Given the fundamental differences in the radiative properties of \CIV and \HeII, the observed \CIV halo suggests that pure photoionization models struggle to explain these results, and that additional mechanisms are required.

Resonance scattering in outflowing gas is frequently invoked as main mechanism to explain the observed \CIV emission from star-forming galaxies and AGN.
Given that the \CIV $\lambda\lambda 1548,1551$ doublet arises from a resonance transition, \CIV photons can undergo multiple scatterings before escaping. In addition, \CIV is produced predominantly through collisional excitation, with a minor contribution from recombination, in warm ionized gas, making it a sensitive tracer of such outflows.
Indeed, powerful galactic winds, with velocities up to $\sim 1000\,\kms$, are commonly observed in star-forming galaxies and AGN. 
However, current studies of radiative transfer of \CIV are insufficient to fully understand its spatial distribution and spectral variations. In this paper, we use a 3D Monte Carlo radiative transfer (RT) code, \texttt{RT-scat} \citep{Chang23,Chang24}, to study \CIV RT as a tracer of warm gas at $T \approx 10^5\,\rm K$.

This paper is organized as follows. 
In \S~\ref{sec:RT_CIV}, we introduce the physics of radiative transfer and describe the details of our 3D Monte Carlo simulations for a spherical scattering medium surrounding a point emission source. In \S~\ref{sec:Result}, we present the simulated spectral profiles of \CIV emission for both a monochromatic source and a Gaussian source, and investigate the doublet ratio of \CIV $\lambda1548$ and $\lambda1551$, and the spatial distribution for the case of a Gaussian source in an outflowing medium. In \S~\ref{sec:Discussion}, we discuss the results of our \CIV radiative transfer simulations. Finally, \S~\ref{sec:Conclusion} summarizes our main conclusions.

\section{Radiative transfer Simulation of C IV Resonance Doublet}\label{sec:RT_CIV}

We use a 3D Monte Carlo radiative transfer code \texttt{RT-scat} \citep{Chang23,Chang24} to study the formation of \CIV resonance doublet as a tracer of warm gas at $T\approx 10^5 \, \rm K$.
In this section, we introduce the atomic physics relevant to \CIV radiative transfer (\S~\ref{sec:cross_section}), the optical depth of \CIV (\S~\ref{sec:optical_depth}), the model geometry and kinematics (\S~\ref{sec:geometry}), and the Monte-Carlo technique (\S~\ref{sec:monte_carlo}).

\subsection{Scattering Cross Section of C~IV}\label{sec:cross_section}

\begin{figure}
    \centering
    \includegraphics[width=\linewidth]{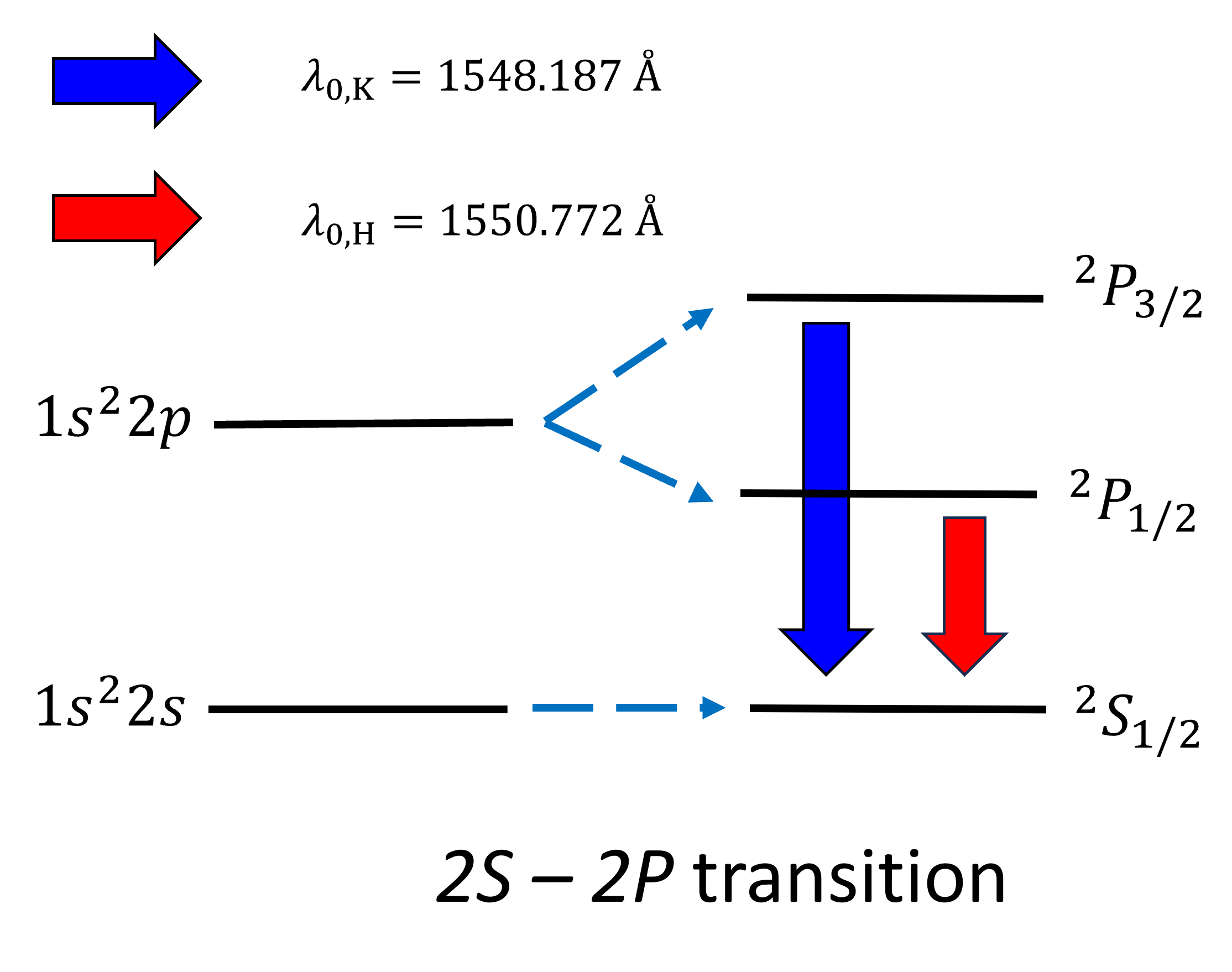}
    \caption{Energy levels of the \CIV resonance doublet at $\lambda\lambda$ 1548, 1551. As a lithium-like ion, the \CIV emission exhibits a resonance doublet arising from $2S-2P$ transitions. The \CIV $\lambda1548$ line (blue), known as the K line, arises from the $2P_{3/2} \rightarrow 2S_{1/2}$ transition with wavelength $\lambda_{0,\rm K} = 1548.187\,$\AA. The \CIV $\lambda1551$ line (red), referred to as the H line, arises from the $2P_{1/2} \rightarrow 2S_{1/2}$ transition with wavelength $\lambda_{0,\rm H} = 1550.772\,$\AA. Due to its larger statistical weight, the K component is intrinsically twice as strong as the H component.}
    \label{fig:Fig_1}
\end{figure}

The \CIV $\lambda\lambda$1548, 1551 resonance doublet arises from the $2S_{1/2}\rightarrow 2P_{1/2,3/2}$ transitions. Hereafter, we refer to the \CIV $\lambda$1548 and $\lambda$1551 components as the `K' and `H' lines, respectively. 
As illustrated in Figure~\ref{fig:Fig_1}, the two transitions have rest wavelengths $\lambda_{0,\rm K} = 1548.187\,$\AA\ and $\lambda_{0,\rm H} = 1550.772\,$\AA, corresponding to the line-center frequencies $\nu_{0,\rm K} = 1.936410\times10^{15}\,\rm s^{-1}$ and $\nu_{0,\rm H} = 1.933182\times10^{15}\,\rm s^{-1}$. Because the K- and H-lines originate from the $2P_{3/2}$ and $2P_{1/2}$ upper levels, they carry different statistical weights ($g = 2J+1$), namely $g_{\rm K} = 4$ and $g_{\rm H} = 2$. 
For this reason, the intrinsic strength of the K line is twice that of the H line, as reflected in the oscillator strengths of $f_{\rm K} = 0.190$ and $f_{\rm H} = 0.0952$, respectively.

The scattering cross section of the \CIV doublet is expressed as a sum of two Voigt--Hjerting functions
\begin{equation}
    \sigma_\nu = \frac{\sqrt{\pi} \, e^2}{m_e c} \left[ \frac{f_{\rm K}}{\Delta \nu_{D,\rm K}} H(x_{\rm K},a_{\rm K}) + \frac{f_{\rm H}}{\Delta \nu_{D,\rm H}} H(x_{\rm H},a_{\rm H}) \right] ,
\label{Eq: cross-section}
\end{equation}
where $m_e$ is the electron mass and $c$ is the speed of light. The Voigt--Hjerting function is defined as
\begin{equation}
    H(x,a) = \frac{a}{\pi} \int^{\infty}_{-\infty} \frac{e^{-y^2}}{(x-y)^2 + a^2} \, dy ,
    \label{eqn_Voigt Hjerting function}
\end{equation}
where the damping parameter $a = \frac{\Gamma}{4\pi \Delta \nu_{\rm D}}$ is the ratio of the damping constant $\Gamma$ to the thermal Doppler width $\Delta \nu_{\rm D}$. The corresponding damping constants are $\Gamma_{\rm K} = 2.65 \times 10^8\,\rm s^{-1}$ and $\Gamma_{\rm H} = 2.64 \times 10^8\,\rm s^{-1}$. The thermal Doppler width is given by $\Delta \nu_{\rm D} = \frac{v_{\rm th}}{c} \nu_0$. Here, $v_{\rm th} = \sqrt{2k_{\rm B}T / m_{\rm CIV}}$ is the thermal velocity, and $k_{\rm B}$ and $m_{\rm CIV}$ denote the Boltzmann constant and the atomic mass of \CIV, respectively. 
We also define the dimensionless frequency parameter as
\begin{equation}
x = \left( \nu - \nu_0 \right) / \Delta \nu_{\rm D},
\label{eqn_x}
\end{equation}
which measures the offset from the line-center frequency $\nu_0$ in units of the Doppler parameter.
For the doublet, the dimensionless parameters $x_{\rm K}$ and $x_{\rm H}$ are defined relative to $\nu_{0,\rm K}$ and $\nu_{0,\rm H}$, with Doppler widths $\Delta \nu_{D,\rm K}$ and $\Delta \nu_{D,\rm H}$, respectively.

\begin{figure}
    \centering
    \includegraphics[width = \linewidth]{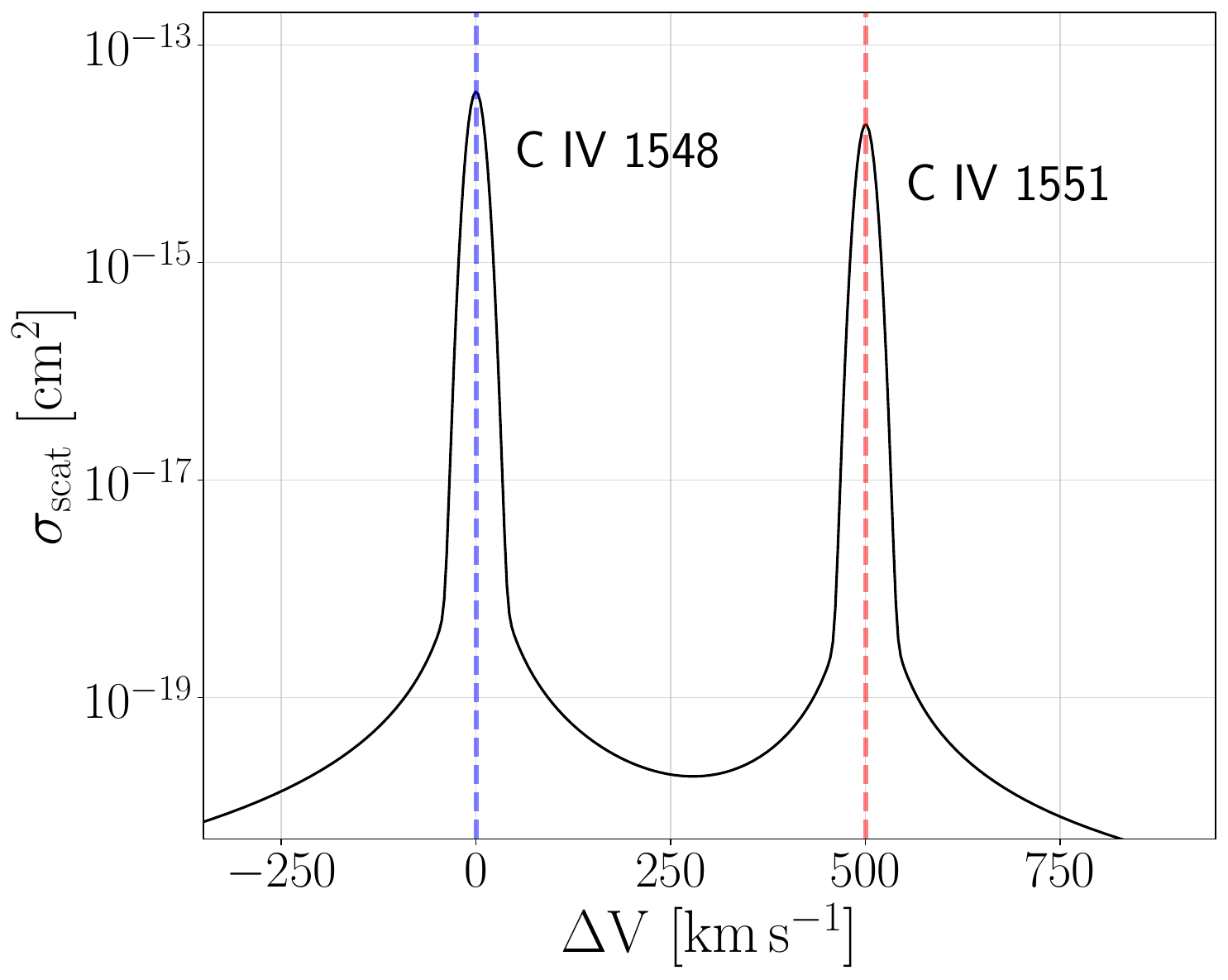}
    \caption{Scattering cross section of the \CIV doublet as a function of the Doppler factor $\Delta V$ defined in Equation~\ref{eqn_velocity_space}. The solid line shows the scattering cross section for the thermal velocity of \CIV at a temperature $T = 10^5 \, \rm K$, corresponding to $v_{\rm th} = 11.8 \rm \, km \, s^{-1}$. The blue vertical dashed line indicates the line center of the K line at $\Delta V = 0\,\kms$, and the red vertical dashed line shows that of the H line. The velocity separation of the two components is $\vsep = 500.9\,\kms$.
    }
    \label{fig:Fig_2}
\end{figure}

Figure~\ref{fig:Fig_2} illustrates the total scattering cross section near the \CIV doublet for a medium with $T = 10^5\,{\rm K}$ using Equation~(\ref{Eq: cross-section}). The horizontal axis represents the Doppler factor $\Delta V$ with respect to the K component, given by
\begin{equation}
\Delta V = \left(\frac{\nu_{0,\rm K}}{\nu} - 1\right) c \simeq -v_{\rm th} x_{\rm K},
\label{eqn_velocity_space}
\end{equation}
where $v_{\rm th} = 11.8\,\kms$ is the thermal speed of carbon at $T = 10^5\,{\rm K}$. As shown in the figure, the velocity separation between the two components with respect to the K component is $\vsep = 500.9\,\kms$, which allows the \CIV doublet to be clearly resolved into two distinct lines. 
The line center cross sections of the K and H components are $3.74 \times 10^{-14}$ and $1.87 \times 10^{-14}\,{\rm cm^2}$, respectively. We note that the cross section decreases sharply, by nearly 4 orders of magnitude, once the absolute frequency parameter reaches $|x_{K,H}| > 3$.

\subsection{Optical Depth at the Line Center}\label{sec:optical_depth}

The optical depth of each component of the \CIV doublet is given by
\begin{equation}
\tau_\nu = \sigma_{\nu} \NCIV = \tauo  H(x,a) ,
\label{eqn_doublet_tau}
\end{equation}
where $\sigma_\nu$ is the scattering cross section from Equation~\ref{Eq: cross-section} and \tauo is the optical depth at the line center.
Since $H(0,a) \simeq 1 $ for $a \ll 1$, \tauo for the K line is given by 
\begin{equation}
\begin{aligned}
    \tau_0 &\equiv \sigma_{\nu_{\rm 0,K}} \NCIV \\
    & \simeq \frac{\sqrt{\pi} e^2}{m_e c} \frac{f_{\rm K}}{\Delta \nu_{\rm D,K}} H(0,a) \NCIV \\ 
    &=\left(\frac{N_{\rm CIV}}{2.67\times10^{13}\,\rm cm^{-2}}\right) \left(\frac{\sigmaR}{11.8\,\rm km\,s^{-1}}\right)^{-1},
\label{Eq:cen_tau}
\end{aligned}
\end{equation}
where $\sigma_{\nu_{\rm 0,K}}$ represents the cross section at the center of the K line and \sigmaR is a velocity dispersion combining thermal and turbulent motions, $\sigmaR = \sqrt{v_{\rm th}^2 + v_{\rm tur}^2}$. Throughout this work, we adopt \sigmaR instead of $v_{\rm th}$. The line-center optical depth for the H line is half that of the K line at the same column density, due to the difference in oscillator strengths.

The optical depth at the line center is an important parameter in spectral line formation, as demonstrated in previous RT simulations for \Lya \citep{Neufeld90,Ahn2000,Dijkstra2006,Laursen2009,Smith2015,Seon2020} and Mg~II \citep{Prochaska2011,Chang24}. In the optically thin regime (\tauo < 1), most photons escape directly, or after a single scattering near the line center. In contrast, when the medium is optically thick (\tauo > 1), most photons are trapped at the line center, undergoing multiple scatterings, and escape only after diffusing in frequency into the wings of the line profile.
\footnote{In a clumpy medium, photons can escape via random walk between the dense clumps through surface scattering even when the covering factor of clumps, which is the average number of clumps in the line of sight, is less than its critical value \citep{Gronke2017}.}
We will investigate the line formation of \CIV for various optical depths in Sections~\ref{sec:Mono} and \ref{sec:Gaussian}.

\subsection{Geometry and C~IV Emission Source}\label{sec:geometry}

 \begin{figure}
    \centering
    \includegraphics[width=\linewidth]{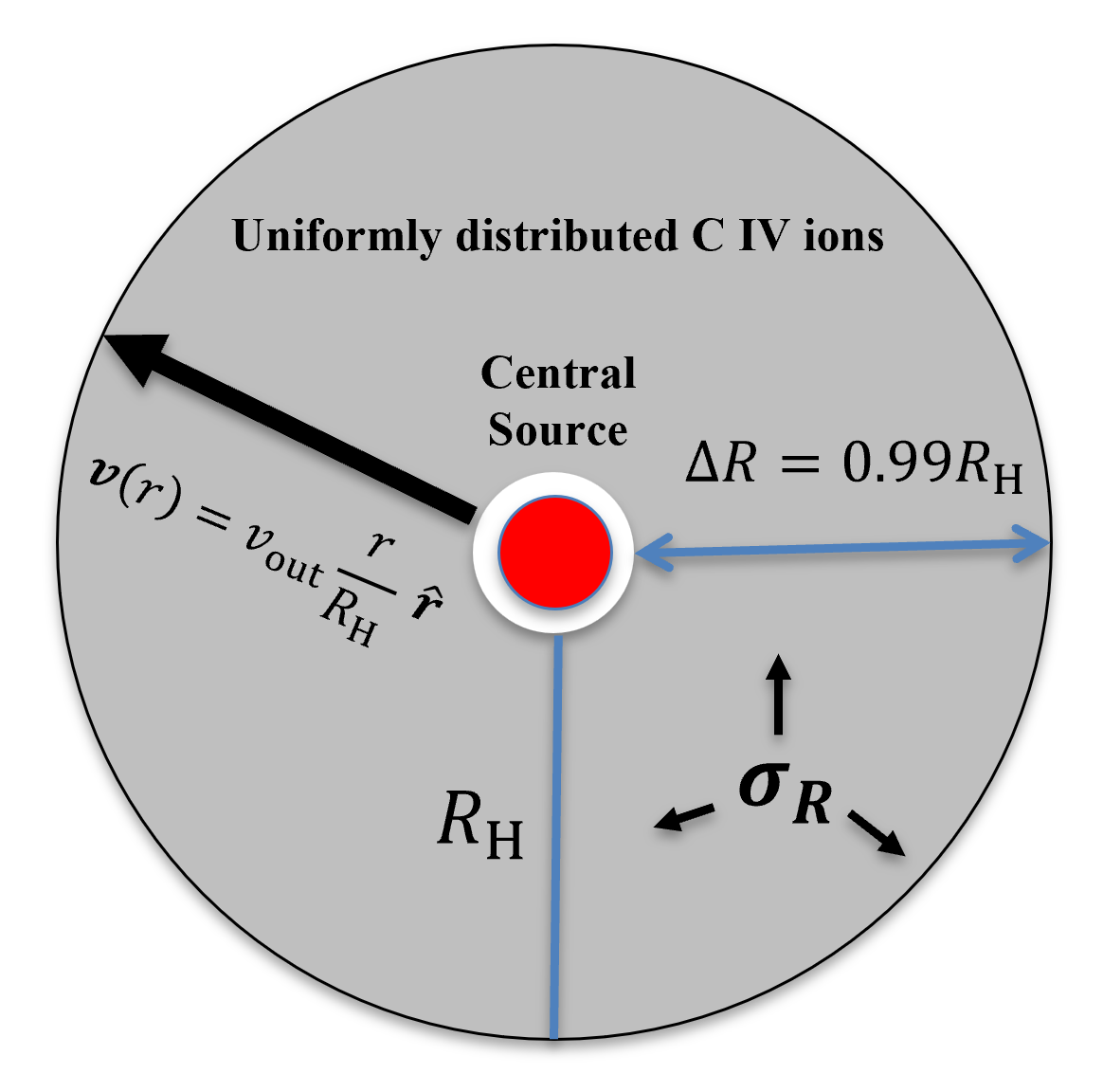}
    \caption{Schematic illustration of the simulation geometry. A spherical \CIV halo of radius $R_{\rm H}$, spanning from $0.01\,R_{\rm H}$ to $R_{\rm H}$ ($\Delta R = 0.99\,R_{\rm H}$), is uniformly filled with \CIV ions (gray region) with a constant number density of $n_{\rm CIV} = \NCIV / \Delta R$. The red circle represents the central source, emitting monochromatic or Gaussian \CIV photons. 
    The scattering medium is characterized by two kinematic components: (i) a radial outflow velocity $v(r) = \vout\,r/R_{\rm H}$, and (ii) velocity dispersion \sigmaR. }
\label{fig:Fig_3}
\end{figure}

To explore the formation of \CIV, we consider a simple geometry composed of a spherical \CIV halo and a central point source emitting the \CIV doublet photons (see Figure~\ref{fig:Fig_3}). 
\CIV ions are uniformly distributed in the spherical halo between radii $0.01\,R_{\rm H}$ and $R_{\rm H}$ ($\Delta R =0.99 R_{\rm H}$), where $R_{\rm H}$ denotes the halo radius. The ions have random velocities characterized by a velocity dispersion \sigmaR. To model an expanding medium, we assume a radial velocity field proportional to the distance from the central source (i.e., $v(r) = \vout\,r/R_{\rm H}$, where $r$ is the distance from the emission source and \vout is the maximum outflow velocity at $R_{\rm H}$). The spherical \CIV halo is characterized by a \CIV column density \NCIV at the outer boundary $R_{\rm H}$, the velocity dispersion \sigmaR, and the maximum outflow velocity \vout.
We explore a range of \NCIV from $10^{12}$ to $10^{17}\,\rm cm^{-2}$, which is broader than typically probed in absorption line spectroscopic studies \citep{Davies2023,Garza2024,Garza2025}. We also consider a wide range of \vout from 0 to $1000\,\kms$, which exceeds the velocity separation between the K- and H-lines ($\vsep \approx 500\,\kms$). When $\vout > \vsep$, the K line photons can be scattered by the H line transition of \CIV in outflowing gas \citep[e.g., as in Mg~II resonance doublet;][]{Chang24,Seon2024}.

We assume a central point source emitting \CIV doublet photons with an intrinsic K-to-H flux ratio of 2, consistent with collisional excitation of \CIV or recombination of C~V. We consider two types of central sources: (i) a monochromatic source for a fundamental study and (ii) a Gaussian source for a realistic case. 
The intrinsic profiles of the K- and H-lines are assumed to be Gaussian, with an intrinsic width $\sigmaint$ spanning 0 to $3000\,\kms$, broad enough to cover sources from star-forming galaxies to AGNs. When $\sigmaint \gtrsim \vsep$, the K- and H-lines are intrinsically mixed and appear as a single broad emission feature that is difficult to resolve into two distinct components, even without radiative transfer effects. We investigate the formation of \CIV line profiles for various intrinsic widths in Section~\ref{sec:spectrum_static}.
In Table~\ref{tab_parameters}, we summarize the parameters that characterize the scattering geometry and \CIV emission, along with their adopted ranges. 

\begin{table}
    \centering
    \renewcommand{\arraystretch}{1.5} 
    \caption{Parameters in our radiative transfer simulation}
    \begin{tabular}{c c c}
        \hline
        \hline
        Parameter & Value/Units & Meaning \\
        \hline
        $\lambda_{0,\rm K}$& 1548.187\AA & Atomic wavelength of K line \\
        \hline
        $\lambda_{0,\rm H}$& 1550.772 \AA & Atomic wavelength of H line \\
        \hline
        $\NCIV$&$ 10^{12-17} \rm cm^{-2}$ & Total \CIV column density \\
        \hline
        $\sigma_R$ & $11.8^{\dagger} - 50 ~ \rm km \, s^{-1} $ & Velocity dispersion of scattering medium \\
        \hline
        $\sigma_{\rm int}$ & $0^*  -  3000  \, \rm km\, s^{-1}$ & Width of intrinsic emission \\
        \hline
        $V_{\rm out}$ & $0 - 1000 ~ \rm km \, s^{-1}$ & Maximum outflow velocity \\
        \hline
    \end{tabular}
    \\
            \footnotesize
        $\dagger$ : thermal speed of \CIV at $T = 10^5~$K,
        $*$: $\sigma_{\rm int} = 0 \kms$ represents the monochromatic case.
    \label{tab_parameters}
\end{table}

\subsection{Monte Carlo Radiative Transfer Simulations}\label{sec:monte_carlo}

Our Monte Carlo simulation is based on the methodology presented in Section~2.5 of \citet{Chang24} and explores the scattering behavior of \CIV doublet, following five steps:

\begin{enumerate}

\item The scattering gas geometry and kinematics are characterized by the physical parameters listed in Table~\ref{tab_parameters}. The physical properties of each cell in a 3D Cartesian grid, with 200 cells per dimension, are assigned based on the position of its center. 

\item We generate a total of $10^{7}$ photons from the central source, assigning each photon a wavelength $\lambda_i$ drawn from a Gaussian profile with width $\sigmaint$ and an isotropically distributed propagation direction $\boldsymbol{\rm k}_i$.

\item The optical depth to the next scattering event is drawn as $\tau_f = -\ln r_u$, where $r_u$ is a uniform random number between 0 and 1. Using the scattering cross section $\sigma_\nu$ defined in Equation~\ref{Eq: cross-section}, the corresponding physical free path is then obtained as $l_f = \tau_f/n_{\rm CIV}\sigma_\nu$, where $n_{\rm CIV}$ is the number density of \CIV ions, $n_{\rm CIV} = \NCIV / \Delta R$. This free path determines the location of the next scattering event.

\item After each scattering event, the direction of the scattered photon $\boldsymbol{\rm k}_s$ and its wavelength $\lambda_s$ are determined by the scattering phase function and the frequency redistribution function, respectively. Since \Lya and \CIV have similar resonance scattering structures, we adopt the \Lya scattering phase function and frequency redistribution function \citep{Chang23}, following the methodology of \citet{Chang24} for the Mg~II resonance doublet.

\item Steps (iii)-(iv) are repeated until the photon escapes the whole grid. Once it escapes, $\lambda_s$ is converted to the observed-frame wavelength, and the escaping photon is collected for the spectrum and surface brightness. The calculation then returns to Step (ii) to generate and propagate the next photon. The Monte Carlo simulation ends once the number of generated photons reaches the specified target total number.

\end{enumerate}

In this study, we neglect polarization or dust effects\footnote{In our framework, dust effects arise from the difference in path lengths between the K- and H-lines, which can be significant for monochromatic sources. However, for broader input spectra, the dust optical depth has little effect on the doublet ratio because the K and H lines experience similar path lengths. In addition, dust in warm gas is expected to be minor.} in the \CIV resonance lines, although \texttt{RT-scat} is capable of modeling both effects, as demonstrated in a previous Mg~II study \citep{Chang24}.

 \begin{figure*}
    \centering
    \includegraphics[width=\linewidth]{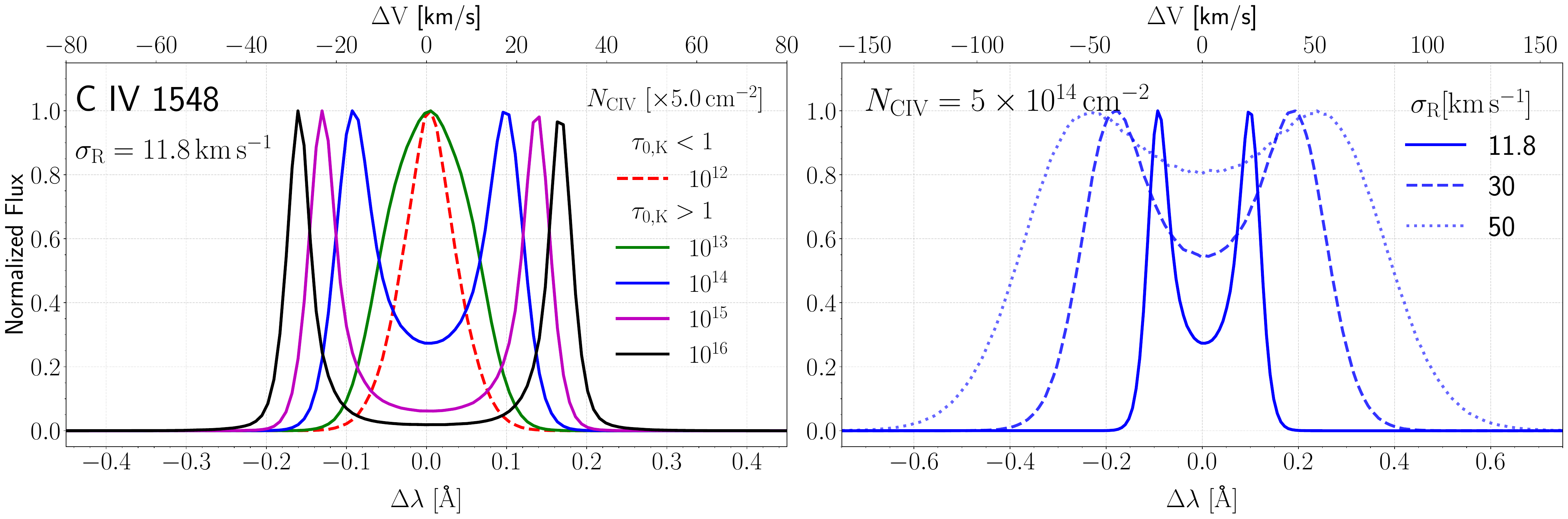}
    \includegraphics[width=\linewidth]{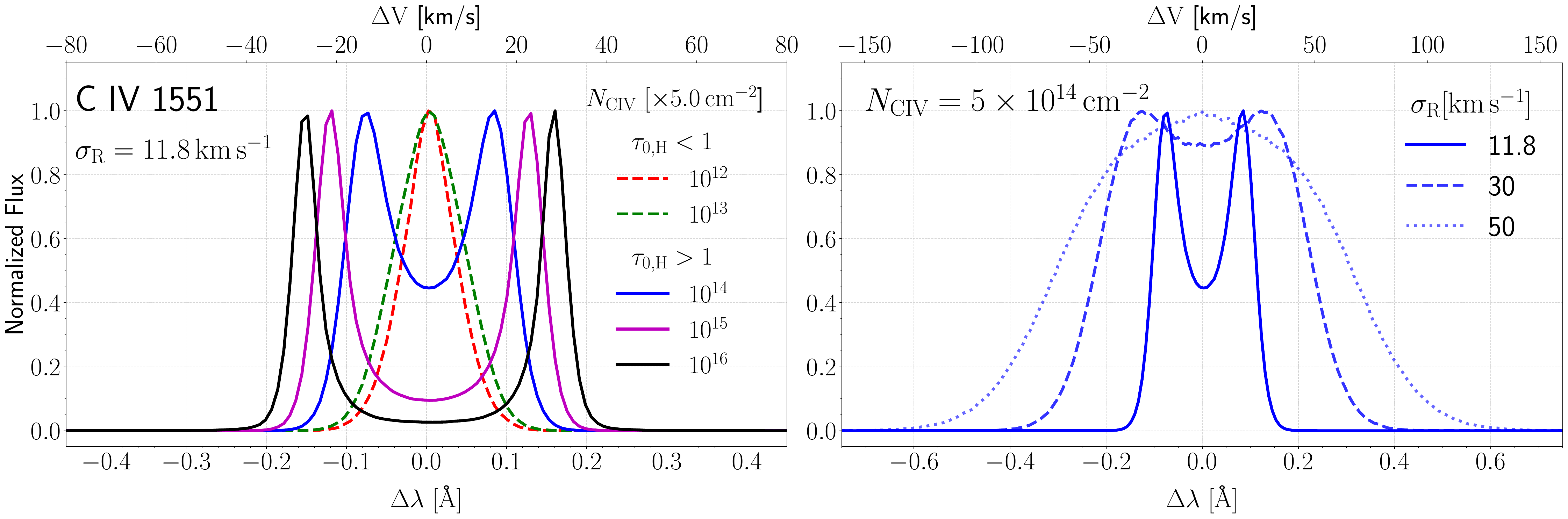}
    \caption{\CIV spectra of scattered photons in a spherical geometry with monochromatic emission and a static medium. The top and bottom panels show the \CIV $\lambda1548$ (K line) and \CIV $\lambda1551$ (H line), respectively, as functions of \NCIV and \sigmaR. The left panels show the spectra for $\NCIV = 5\times10^{12}$--$10^{16}\,\unitNHI$ at $\sigmaR = 11.8\,\kms$. $\tau_{\rm 0,K}$ and $\tau_{\rm 0,H}$ denote the optical depths at the line center of the K- and H-lines (Equation~\ref{Eq:cen_tau}). The dashed lines indicate the optically thin regime ($\tauo < 1$), while the solid lines show the optically thick regime ($\tauo > 1$). The right panels illustrate the spectra for $\sigmaR = 11.8$, 30, and $50\,\kms$ at $\NCIV = 5.0\times10^{14}\,\unitNHI$. All spectra are normalized to a peak value of unity.
    }
    \label{fig:Fig_4}
\end{figure*}

\section{Result}\label{sec:Result}

In this section, we present the results of our \CIV radiative transfer modeling. Section~\ref{sec:Mono} describes the formation of the \CIV spectrum in a static medium with a monochromatic source. Section~\ref{sec:Gaussian} examines the \CIV spectrum for a Gaussian source in two cases: (i) a static medium and (ii) an outflowing medium. For the outflowing medium case, we also explore the \CIV doublet ratio as a diagnostic of fast outflows. In Section~\ref{sec:Spatial_distribution}, we investigate the spatial distribution of \CIV emission, including photoionization and scattering effects.

\subsection{Line formation of C IV--Monochromatic source}\label{sec:Mono}
To study the fundamental effects of radiative transfer on the \CIV doublet, we begin with idealized conditions: a static medium (\vout = 0\,\kms) and a monochromatic source (\sigmaint = 0\,\kms).
Figure~\ref{fig:Fig_4} presents \CIV spectra of scattered photons for different \CIV column densities \NCIV and velocity dispersions \sigmaR.

In the left panels of Figure~\ref{fig:Fig_4}, the spectrum broadens as $\NCIV$ increases. At $\NCIV = 5\times10^{12}\,\unitNHI$, the K and H components of the \CIV doublet are optically thin at line center, and each scattered component exhibits a single Gaussian profile. In this regime, single scattering is dominant, and the line width is primarily determined by the velocity dispersion \sigmaR. In contrast, at $\NCIV \geq 10^{14}\,\unitNHI$, the line-center optical depth exceeds 10 for both lines. As a result, scattered photons undergo multiple scattering and produce clear double-peaked profiles with a central dip.

In the right panels of Figure~\ref{fig:Fig_4}, the K and H line profiles also broaden with increasing \sigmaR, because the scale of frequency shift in each scattering depends on the velocity dispersion (see Equation~\ref{eqn_x}). Additionally, the central suppression becomes less pronounced because the line center optical depth, \tauo, decreases as \sigmaR increases (see Equation~\ref{Eq:cen_tau}).

The K and H line profiles in Figure~\ref{fig:Fig_4} differ even at fixed \sigmaR for \NCIV = $5\times10^{14}\,\unitNHI$, owing to their different scattering cross sections.
However, at higher \NCIV ($> 10^{15} \unitNHI$, i.e., \tauo > 100), the spectral profile of the K line becomes similar to that of the H line.
For example, in the left panels of Figure~\ref{fig:Fig_4}, the K line profile at \NCIV = $5\times10^{16} \unitNHI$ in the top panel closely resembles the H line profile in the bottom panel.

As shown in Figure~\ref{fig:Fig_4}, the scattered \CIV spectrum broadens with increasing \NCIV and \sigmaR. The line profile also changes from a single-peaked to a double-peaked as \NCIV increases, while the double-peaked structure becomes less pronounced as \sigmaR increases. Given that \tauo is jointly determined by \NCIV and \sigmaR (see Equation~\ref{Eq:cen_tau}), the optical depth at the line center plays a key role in shaping the line profile.
As described in Section~\ref{sec:optical_depth}, scattered photons escape directly at the line center in the optically thin regime ($\tauo < 1$), whereas in the optically thick regime ($\tauo > 1$), they are trapped near the line center and undergo multiple scatterings. 
This indicates that multiple scattering is the primary mechanism driving the variation of the \CIV spectrum.
In the following section, we investigate the impact of multiple scattering on line broadening.

\subsubsection{Line Broadening}
\label{sec:line_broadening}

\begin{figure}
    \centering
    \includegraphics[width = \linewidth]{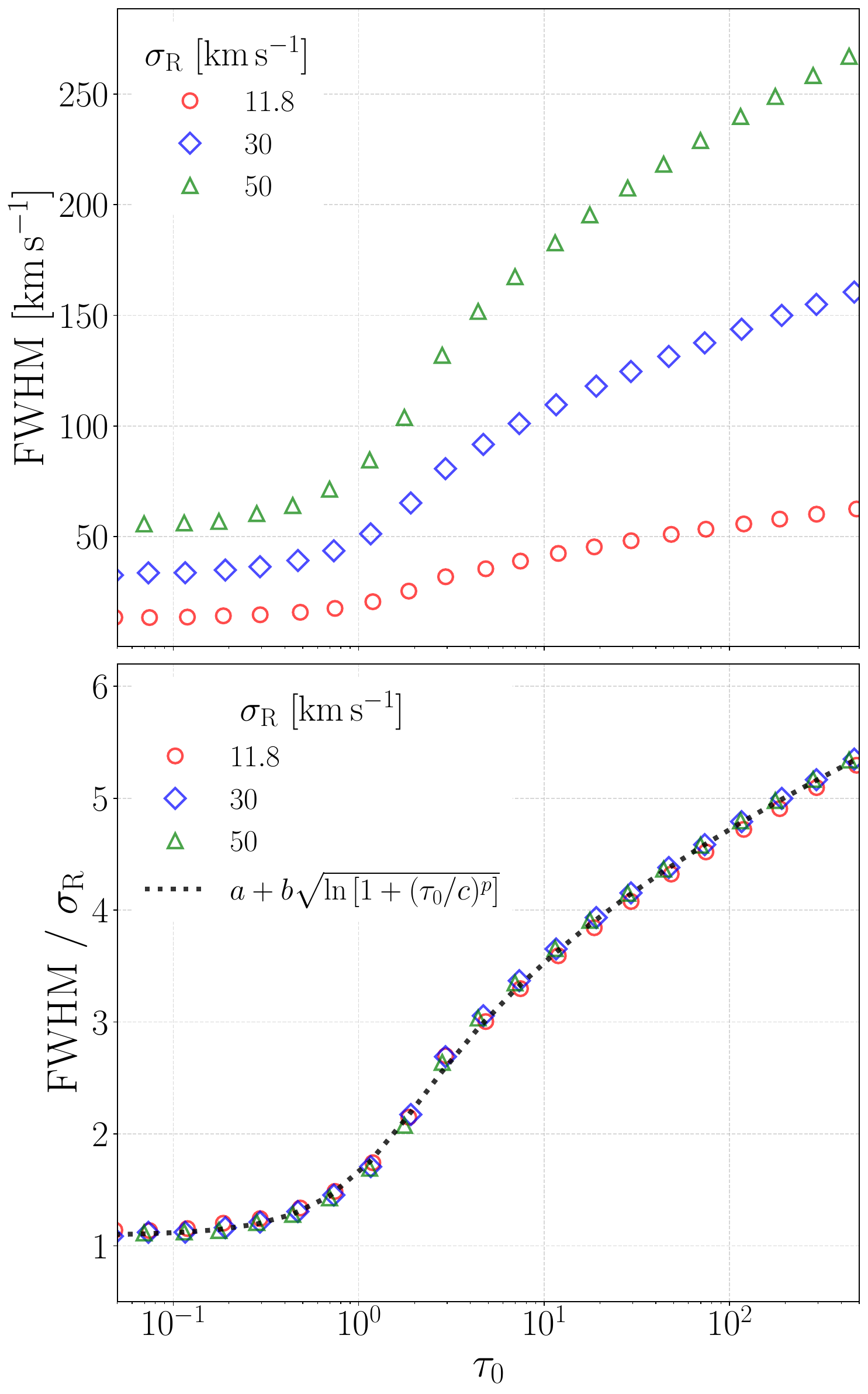}
    \caption{Full width at half maximum (FWHM; top) and the ratio of the FWHM to the velocity dispersion \sigmaR (bottom) of the \CIV K line as a function of the optical depth at the K line center \tauo. 
    The FWHM is measured differently depending on \tauo: in the standard manner for $\tauo < 1$, and as the width between the outermost half-maximum points of the double-peaked profile for $\tauo > 1$.
    In both panels, the colored symbols denote \sigmaR = 11.8 (red circles), 30 (blue diamonds), and 50\,\kms (green triangles). In the bottom panel, the dotted line represents the empirical relation (Equation~\ref{eqn_empirical}).
    }
    \label{fig:Fig_5}
\end{figure}

To investigate line broadening, we measure the full width at half maximum (FWHM) of the scattered \CIV spectrum as a function of \tauo.
Since the shape of the \CIV spectrum depends on \tauo, we adopt different definitions of the FWHM depending on the optical depth.
For \tauo < 1, the scattered \CIV profile is well described by a Gaussian, and the FWHM is measured in the standard manner.
For \tauo > 1, however, the profile exhibits a double-peaked structure, making the standard definition of the FWHM ambiguous.
In this regime, we define the FWHM as the width between the outermost points at half the peak intensity.
In Figure~\ref{fig:Fig_5}, we present the FWHM for different velocity dispersions, $\sigmaR = 11.8$, 30, and 50\,\kms. In addition, we present the ratio FWHM$/\sigmaR$ to highlight the effect of line broadening. Both $\tauo$ and the FWHM are calculated using the \CIV K line.

The top panel of Figure~\ref{fig:Fig_5} shows that the FWHM increases with both \tauo and \sigmaR. 
In the optically thin regime ($\tauo < 1$), where single scattering dominates, the FWHM remains nearly constant.  
In contrast, in the optically thick regime ($\tauo > 1$), multiple scattering becomes dominant and the FWHM increases with \tauo. The FWHM also increases with \sigmaR, since the frequency shift per scattering is proportional to the velocity dispersion of the medium.

The bottom panel of Figure~\ref{fig:Fig_5} shows the FWHM normalized by \sigmaR, highlighting the relative contribution of multiple scatterings to line broadening. 
The dependence of FWHM/\sigmaR on \tauo is nearly identical for different values of \sigmaR. 
When \tauo < 1, the ratio remains nearly constant and is smaller than that for a Gaussian, for which ${\rm FWHM}/\sigma = 2\sqrt{2\ln2}$ $\simeq$ 2.35. This suggests that, in the optically thin regime, the line is not broadened to the full extent expected from the velocity dispersion of the medium.
The ratio increases with \tauo and is well described by the empirical relation
\begin{equation}
\frac{\rm FWHM}{\sigmaR} = a + b \sqrt{\ln\left[1 + \left(\tauo/c\right)^{p}\right]},
\label{eqn_empirical}
\end{equation}
where $(a, b, c, p) = (1.092, 1.061, 1.505, 2.788)$ are the best-fit parameters.
In the optically thin limit ($\tauo \ll 1$), this relation approaches a constant value ($a$). In contrast, in the optically thick limit ($\tauo \gg 1$), it scales as $\sim \sqrt{\ln\tauo}$, in analogy with the saturated regime of the curve of growth, where the equivalent width also grows as $\sqrt{\ln\tauo}$ \citep[e.g.,][]{Draine_book2011}.
Although \sigmaR sets the characteristic width, this relation indicates that the degree of line broadening is mainly governed by \tauo. 
In other words, $\tauo$, which reflects the importance of multiple scattering, plays a crucial role in shaping the \CIV spectrum.

Line broadening caused by multiple scattering has been reported in several observations. 
For example, \cite{Berg2019B} found that the \CIV profile is broader than non-resonance lines such as \HeII and \OIII in two local star-forming galaxies. They showed that the \CIV emission is approximately twice as broad as the non-resonance lines, consistent with $\tauo \sim 1\text{--}10$ in our simulations.
This suggests that the observed \CIV spectrum has undergone significant multiple scattering.
In the following section, we compare our numerical results with analytic solutions to quantify the effects of multiple scattering.

\subsubsection{Numerical solution compared with analytic solution}
\label{sec:analytic_sol}

\begin{figure}
    \centering
    \includegraphics[width = \linewidth]{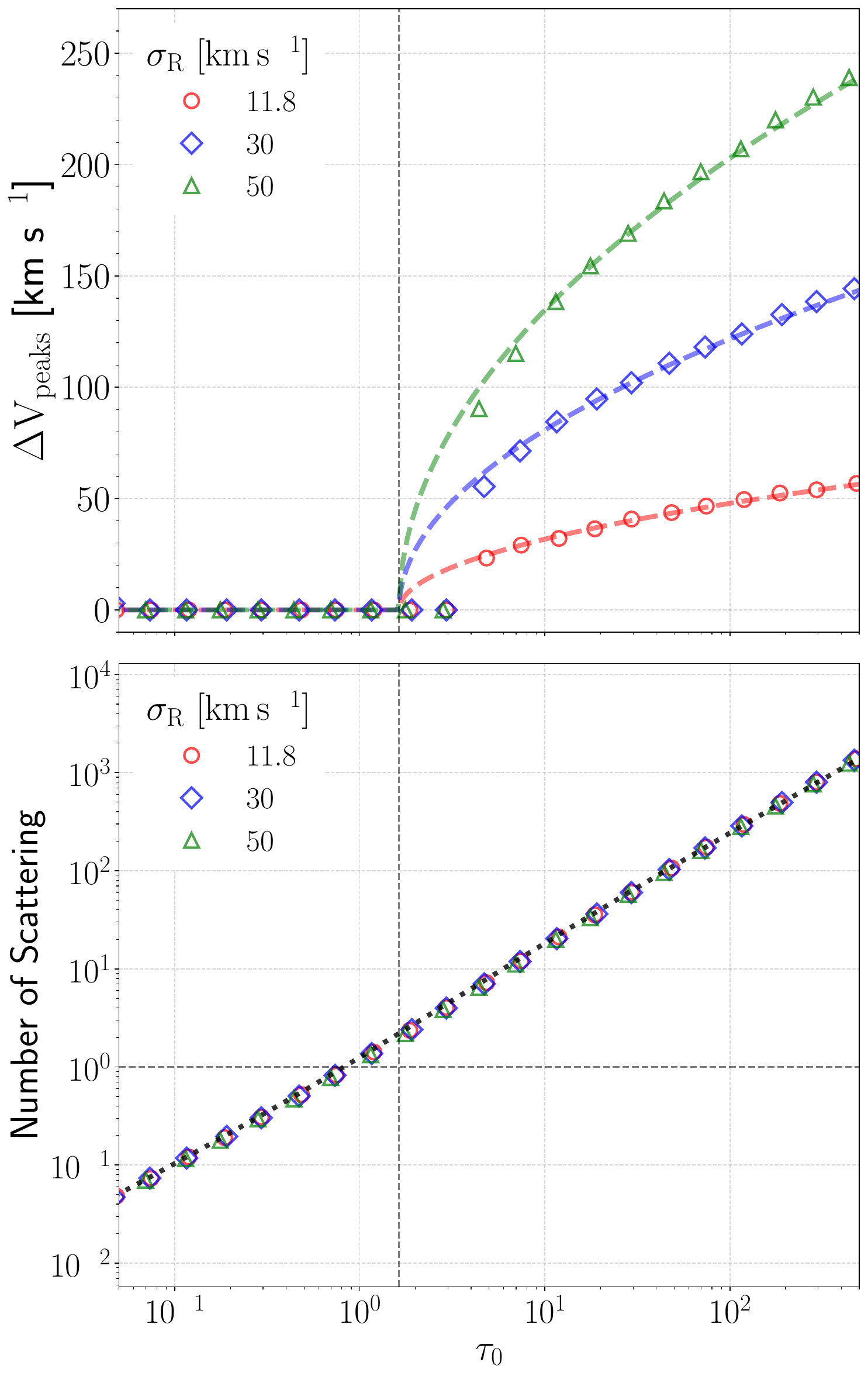}
    \caption{Peak separation of the double-peaked spectra $\Delta V_{\rm peaks}$ (top) and the mean number of scatterings $\langle N_{\rm scat} \rangle$ (bottom). In both panels, the colored symbols denote numerical solutions from \texttt{RT-scat} for the velocity dispersion of the scattering medium: $\sigmaR = 11.8$ (red circles), 30 (blue diamonds), and $50\,\kms$ (green triangles). The vertical dashed lines indicate $\tauo = 1$ at the K line center. In the top panel, colored dashed lines show the analytic solutions following Equation~\ref{eqn_peaks_sep}.
    In the bottom panel, the dotted line presents the analytic mean number of scatterings following Equation~\ref{eqn_NoS_bridge} and the dashed horizontal line marks $\langle N_{\rm scat} \rangle = 1$.}
    \label{fig:Fig_6}
\end{figure}

As discussed in Section~\ref{sec:line_broadening}, the emergent line profile primarily depends on the optical depth. 
To quantify the effects of multiple scattering, we measure the peak separation $\Delta V_{\rm peaks}$ and the mean number of scatterings $\langle N_{\rm scat} \rangle$ of the \CIV K line in our RT simulations and compare them with analytic estimates \citep[e.g.,][]{Oster62,Adams72,Neufeld90}.

To derive analytical estimates for $\Delta V_{\rm peaks}$ and $\langle N_{\rm scat} \rangle$, we first define the threshold frequency beyond which photons can escape the medium after diffusing in frequency space through multiple scatterings.
Spectral peaks naturally emerge near the frequency at which the medium becomes optically thin.
We designate this as the escape frequency, $\nu_{\rm esc}$, with its corresponding dimensionless frequency parameter denoted as $x_{\rm esc}$.
Photons satisfying $|x| > x_{\rm esc}$ can escape the system either directly or after undergoing only $\approx 1$--$2$ scatterings. 

Photons diffuse simultaneously in both real space and frequency space, as illustrated in \S~\ref{sec:monte_carlo}, while in moderately optically thick media ($1\lesssim\tau_0\lesssim 10^4$), the frequency-space diffusion remains within the Gaussian core regime \citep{Draine_book2011}.
Just prior to their last scattering event, scattered photons reach a frequency parameter of $x \approx x_{\rm esc}$ and establish a characteristic spatial distribution within the sphere.
We do not explicitly model the spatial distribution of the last-scattering points. However, by definition, at the location of the last scattering, the medium is effectively optically thin to photons with the escape frequency parameter $x\approx x_{\rm esc}$, even if the system remains optically thick to line-center photons ($x=0$). We can infer $x_{\rm esc}$ from the condition $N_{\rm scat}(x_{\rm esc})\approx 1$, together with the expressions for the mean number of scatterings, $N_{\rm scat}(x_{\rm esc})$, for two limiting spatial distributions of monochromatic photons in an optically thin medium. For a central point source emitting photons at a frequency parameter $x_{\rm esc}$, the mean number of scatterings is given by
$N_{\rm scat}(x_{\rm esc}) = \tau(x_{\rm esc})$ \citep{Rybicki1986}, where $\tau(x_{\rm esc}) = \tauo H(x_{\rm esc},a) \simeq \tau_0 e^{-x_{\rm esc}^2}$ is the optical depth from the center of the sphere to its outer boundary in the Gaussian core regime.
For a spatially uniform source distribution, the corresponding mean number of scatterings is $N_{\rm scat}(x_{\rm esc}) = (3/4)\tau(x_{\rm esc}) \simeq (3/4)\tau_0 e^{-x_{\rm esc}^2}$ \citep{Osterbrock_Ferland_book2006}. These two limiting cases provide a basis for estimating the mean number of scatterings without explicitly specifying the spatial distribution of the last-scattering points.
Specifically, they motivate expressing the generalized mean number of scatterings for an arbitrary spatial distribution of photons with $x\approx x_{\rm esc}$ as $N_{\rm scat}(x_{\rm esc}) = (1/f_*)\tau_0 e^{-x_{\rm esc}^2}$, where $f_* \approx 1$--$2$ is a geometric factor that accounts for the spatial distribution of photons at the escape frequency $x_{\rm esc}$. Based on our simulations of the \CIV emission line, we find that $f_* = 1.63$ provides an appropriate description of our system.
For photons at their last-scattering points, we set the mean number of scatterings to $N_{\rm scat}(x_{\rm esc})\approx 1$, which yields the following estimate for the escape frequency:
\begin{equation}
    x_{\rm esc} = 
    \begin{cases}
        0  & (\tau_{0} \lesssim 1), \\[4pt]
        \sqrt{\ln(\tauo/f_*}) & (1 \lesssim \tau_{0} \lesssim 10^4;\ 0<x_{\rm esc}\lesssim 3).
    \end{cases}
    \label{eqn_x_esc}
\end{equation}
In the optically thin regime, we set $x_{\rm esc} = 0$ because no double-peaked profile is produced.
The second branch applies when a double-peak profile develops and $x_{\rm esc}$ remains within the Doppler-core regime, $0<x_{\rm esc}\lesssim 3$. The corresponding range of $\tau_{0}$ follows from the above relation, $(\tau_0/f_*) e^{-x_{\rm esc}^2}=1$, which gives $\tau_0=f_* e^{x_{\rm esc}^2}$. For $f_*=1.63$ and $0<x_{\rm esc}\lesssim3$, this corresponds approximately to $1\lesssim\tau_0\lesssim10^4$.
Because the profile is symmetric about line center, the separation between the two peaks is given by
\begin{equation}
\Delta V_{\rm peaks} = 2\,\sigmaR \lvert x_{\rm esc} \rvert.
\label{eqn_peaks_sep}
\end{equation}

To derive the mean ``total'' number of scatterings $\langle N_{\rm scat} \rangle$, it is necessary to account for all scatterings experienced by a photon from its initial emission at $x=0$ to its eventual escape at $x\approx x_{\rm esc}$.
Due to the thermal motions, the frequency of a photon after each scattering follows the probability distribution
\begin{equation}
    p(x) = \frac{1}{\sqrt{\pi}} e^{-x^2},
\end{equation}
where $e^{-x^2}$ term is from 1D Maxwell Boltzmann distribution \citep{Oster62}.
A photon escapes once it is re-emitted at a frequency beyond the escape frequency (i.e., $ \lvert x \rvert \geq \lvert x_{\rm esc} \rvert$), which occurs with probability
\begin{equation}
    w = 2\int_{x_{\rm esc}}^{\infty} p(x)dx=1 - \frac{2}{\sqrt{\pi}}\int_0^{\lvert x_{\rm esc}\rvert} e^{-x^2}\, dx = \mathrm{erfc}(\lvert x_{\rm esc}\rvert).
    \label{eqn_escape_prob}
\end{equation}
Then, the mean number of scatterings before escape is
\begin{equation}
    \langle N_{\rm scat} \rangle = w^{-1} = \left[\mathrm{erfc}(\lvert x_{\rm esc}\rvert)\right]^{-1}.
    \label{eqn_analytic_NoS}
\end{equation}
In the two limiting regimes, this expression reduces to
\begin{equation}
    \langle N_{\rm scat} \rangle \approx
    \begin{cases}
        \tauo & (\tauo < 1), \\[4pt]
        \left(\frac{e^{-x_{\rm esc}^2}}{{\sqrt{\pi}\, x_{\rm esc}}}\right)^{-1} 
        = \tauo \sqrt{\dfrac{\pi}{f_*^2}\ln\left(\dfrac{\tauo}{f_*}\right)} & (1 < \tauo \lesssim 10^4),
    \end{cases}
    \label{eqn_NoS_limits}
\end{equation}
where the optically thin result follows from the scattering probability, $1 - e^{-\tauo} \approx \tauo$, and the optically thick result is obtained using the asymptotic expansion $\mathrm{erfc}(x) \approx e^{-x^2}/(\sqrt{\pi}\,x)$.
These two limiting cases can be combined into
\begin{equation}
    \langle N_{\rm scat} \rangle \approx \tauo \left[ 1 + \frac{\pi}{f_*^2}\ln\left( 1 + \frac{\tauo}{f_*} \right) \right]^{1/2}.
    \label{eqn_NoS_bridge}
\end{equation}

Figure~\ref{fig:Fig_6} compares the RT simulation results for $\Delta V_{\rm peaks}$ and $\langle N_{\rm scat} \rangle$ with their analytic estimates as functions of the line center optical depth, $\tauo$.
The figure demonstrates that both quantities increase with $\tauo$.
In the optically thin regime, $\Delta V_{\rm peaks}$ is consistent with zero and $\langle N_{\rm scat} \rangle$ is less than unity. 
This is because most photons escape directly or after a single scattering near line center, implying $x_{\rm esc} = 0$ (Equation~\ref{eqn_x_esc}). The line profile is therefore expected to be single-peaked, with zero peak separation, and $\langle N_{\rm scat} \rangle$ is determined by the scattering probability, $1 - e^{-\tauo} \approx \tauo$ (Equation~\ref{eqn_NoS_limits}).
In contrast, in the optically thick regime, photons are trapped near line center and escape only after undergoing frequency diffusion through multiple scatterings, producing a double-peaked profile and significant line broadening. 
In both regimes, the numerical results are in good agreement with the analytic solutions.

Observationally, \CIV column densities in astrophysical systems lie in the range $\NCIV \sim 10^{13\text{--}15}\,\unitNHI$ \citep[e.g.,][]{codoreanu2018,Rudie2019,Davies2023}, corresponding to $\tauo \sim 10^{-1}$--$10^{1}$.
Our simulations show that $\langle N_{\rm scat} \rangle$ spans $\sim 10^{-1}\text{--}10^{2}$ over this range.
This indicates that multiple scattering can significantly shape the emergent spectrum, especially at $\tauo \gtrsim1$ (i.e., $\NCIV \gtrsim 10^{14}\,\unitNHI$), where $\langle N_{\rm scat} \rangle$ exceeds unity. Therefore, multiple scattering should be taken into account when modeling \CIV line profiles.

In summary, we explored the effects of resonance scattering on \CIV line formation in an idealized case with a monochromatic source and a static medium. We found that multiple scattering plays an important role, producing a double-peaked profile with a central dip and significant line broadening. In the following section, we examine \CIV line formation in a more realistic case using a Gaussian source.

\subsection{The formation of the C IV spectrum--Gaussian Source}
\label{sec:Gaussian}

In this section, we explore the formation of the \CIV doublet for a Gaussian source ($\sigmaint > 0\,\kms$) as a function of the intrinsic line width, $\sigmaint$, for two types of scattering media: static and outflowing media. We adopt a fixed velocity dispersion $\sigmaR = 11.8\,\kms$, corresponding to the thermal speed at $T = 10^5\,$K. In addition, since the velocity separation between the K and H components is $\vsep \simeq 500\,\kms$, we define the K and H line regions as the velocity ranges within $\Delta V = 250\,\kms$ from each line center.

\subsubsection{\CIV spectra from a static medium }
\label{sec:spectrum_static}

\begin{figure}
    \centering
    \includegraphics[width=\linewidth]{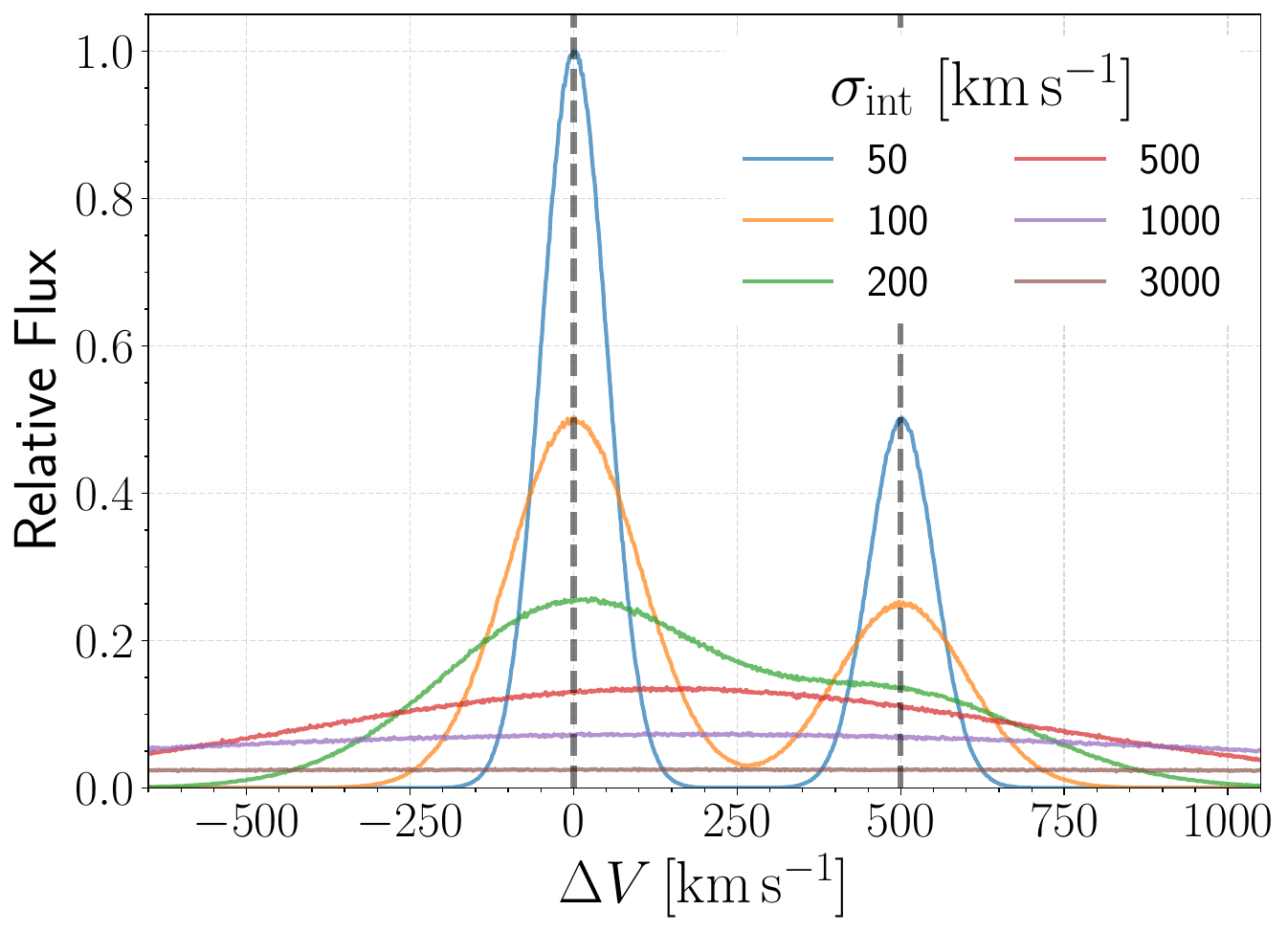}
    \caption{Intrinsic \CIV spectra for various intrinsic emission widths \sigmaint. 
    The colored solid lines represent the \CIV spectra for $\sigmaint = 50$ (blue), 100 (orange), 200 (green), 500 (red), 1000 (purple), and 3000 (brown) $\kms$.  
    The dashed vertical lines indicate the line centers of the K and H components. All spectra are normalized at the velocity corresponding to the peak of the \CIV profile for $\sigmaint = 50\,\kms$.}
    \label{fig:Fig_7}
\end{figure}

As shown in \S~\ref{sec:geometry}, we assume a broad range of an intrinsic width \sigmaint from 0 to 3000\,\kms (cf. Table~\ref{tab_parameters}).
Unlike a monochromatic source (\sigmaint = 0\,\kms), the doublet components of a Gaussian source can be intrinsically blended when $\sigmaint$ is sufficiently broad relative to their separation, $\vsep \simeq 500\,\kms$, even in the absence of radiative transfer effects. To isolate this intrinsic effect, we examine the \CIV doublet spectra in a static, optically thin medium.

Figure~\ref{fig:Fig_7} shows \CIV doublet spectra for $\sigmaint = 50$, 100, 200, 500, 1000, and $3000\,\kms$. For narrow intrinsic widths (i.e., $\sigmaint \leq 100\,\kms$), the K and H components are clearly resolved, as is commonly observed in the spectra of stars and star-forming galaxies.
As $\sigmaint$ increases beyond $100\,\kms$, the two lines begin to overlap. When $\sigmaint \gtrsim \vsep \simeq 500\,\kms$, the two components become fully blended, making the doublet structure indistinguishable. 
This motivates our choice of $\sigmaint = 500\,\kms$ as a representative AGN case in the subsequent simulations, even though observed AGN typically exhibit larger intrinsic widths.

\subsubsection{\CIV spectra from an outflowing medium}
\label{sec:spectrum_outflow}

\begin{figure*}
    \centering
    \includegraphics[width=\linewidth]{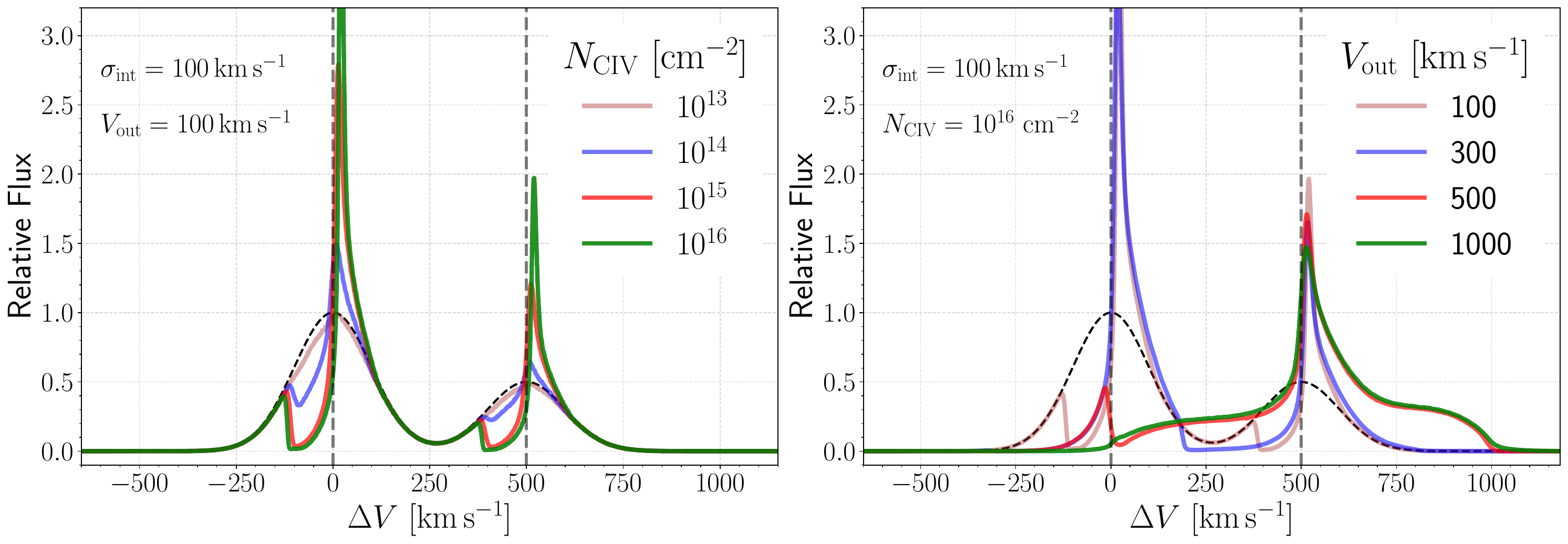} 
    \caption{\CIV spectra for various \CIV column densities $\NCIV$ and outflow velocities $\vout$. The velocity dispersion \sigmaR is fixed at 11.8\,\kms. 
    The left and right panels show the \CIV spectra at \sigmaint = 100\,\kms, where the K- and H-lines are clearly separated.
    The left panel presents the \CIV spectra for $\NCIV = 10^{13-16}\,\unitNHI$ at $\vout = 100\,\kms$. The right panel shows the spectra for $\vout = 100$, 300, 500, and $1000\,\kms$ at $\NCIV = 10^{16}\,\unitNHI$.
    The dashed black line represents the intrinsic Gaussian emission profile for \sigmaint = 100\,\kms. The dashed vertical lines indicate the line centers of the K- and H-lines at $\Delta V = 0$ and 500\,\kms, respectively. All spectra are normalized to the peak of the intrinsic profile.
     }
    \label{fig:Fig_8}
\end{figure*}

\begin{figure*}
    \centering
    \includegraphics[width=\linewidth]{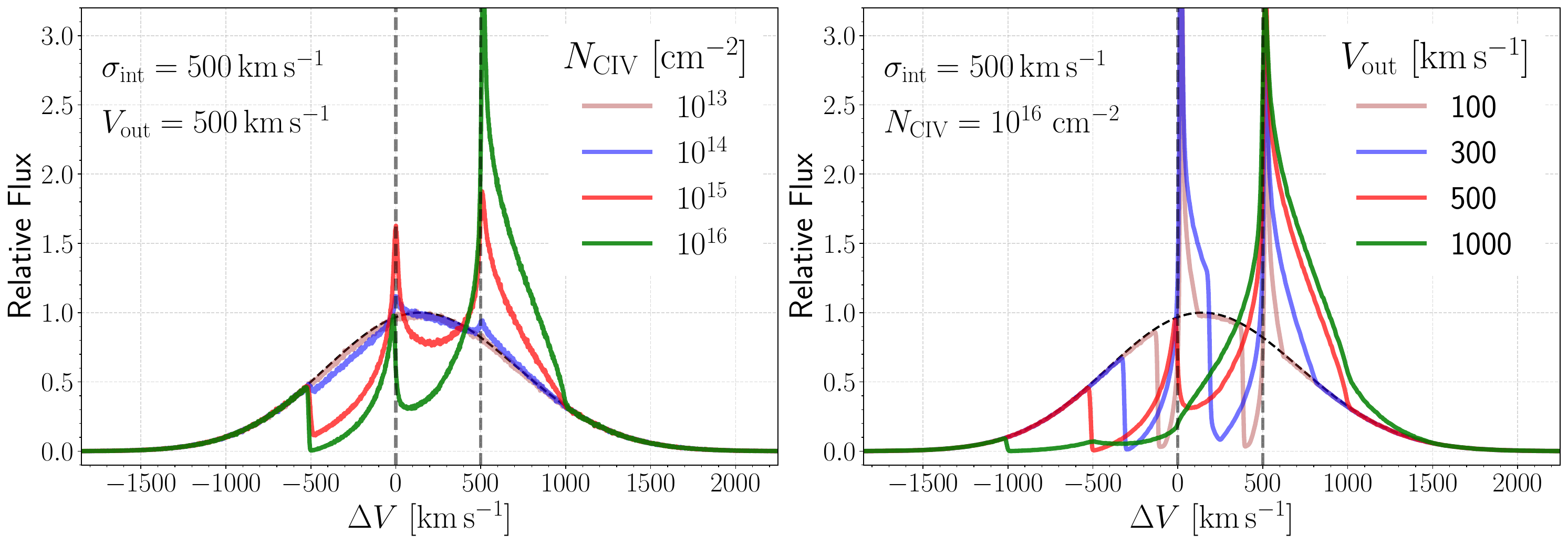}
    \caption{\CIV spectra for various \CIV column densities $\NCIV$ and outflow velocities $\vout$.
    The left and right panels present the \CIV spectra for $\sigmaint = 500\,\kms$, where the K- and H-lines are blended.
    The left panel shows the \CIV spectra for $\NCIV = 10^{13}$--$10^{16}\,\unitNHI$ at $\vout = 500\,\kms$, while the right panel shows the spectra for $\vout = 100$, 300, 500, and $1000\,\kms$ at $\NCIV = 10^{16}\,\unitNHI$.
    The dashed black line represents the intrinsic Gaussian emission profile for $\sigmaint = 500\,\kms$.
    The vertical lines indicate the line centers of the K and H transitions, following the same convention as in Figure~\ref{fig:Fig_8}.}
    \label{fig:Fig_9}
\end{figure*}

We now discuss how resonance scattering in an outflowing medium modifies the intrinsically broad, or even blended, \CIV spectrum.
Figures~\ref{fig:Fig_8} and \ref{fig:Fig_9} show the \CIV spectra for two Gaussian sources in an outflowing medium, one with clearly separated K- and H-lines (\sigmaint = 100\,\kms) and the other with fully blended lines (\sigmaint = 500\,\kms), as functions of \NCIV and \vout. Here, we use \NCIV instead of optical depth since \sigmaR is fixed at 11.8\,\kms.

The left panel of Figure~\ref{fig:Fig_8} shows the \CIV spectra for $\NCIV = 10^{13}$--$10^{16}\,\unitNHI$ at an outflow velocity of $\vout = 100\,\kms$. When $\NCIV = 10^{13}\,\rm cm^{-2}$, the \CIV spectrum retains a symmetric Gaussian profile similar to the intrinsic one, as most photons escape without scattering or after only a single scattering. For $\NCIV \geq 10^{14}\,\unitNHI$, where multiple scattering becomes significant, both the K- and H-lines exhibit profiles characterized by a blue-side absorption-like feature and an enhanced red peak (i.e., a P-Cygni profile). This occurs because outflows redistribute photons from the blue side to the red side through scattering.

The right panel of Figure~\ref{fig:Fig_8} presents the \CIV spectra for $\vout = 100$, 300, 500, and $1000\,\kms$ at $\NCIV = 10^{16}\,\rm cm^{-2}$. When the outflow velocity is smaller than the velocity separation between the K- and H-lines (i.e., $\vout < \vsep \approx 500\,\kms$), both lines show a blue-side absorption-like feature extending from $-\vout$ to $0\,\kms$. At $\vout = 300\,\kms$, the absorption feature of the H line encroaches on the K line region since $\vout$ exceeds the $250\,\kms$ criterion for the K and H line regions, causing a steep drop near the red wing of the K line.

When $\vout \geq \vsep$, the \CIV spectrum is dramatically modified. The intrinsic emission near the K line is suppressed, and the emergent spectrum exhibits a broad wing around the H line, as seen in the case of $\vout = 1000\,\kms$. In this regime, strong outflows allow the H line transition to scatter K line photons, redistributing them around the H line. This result implies that outflows play a key role in mixing K and H line photons, which may account for the observed dominance of the \CIV $\lambda$1551 component in star-forming galaxies \citep{Berg2019A,Schaerer2022,Topping2024}.

Figure~\ref{fig:Fig_9} presents the \CIV spectra with $\sigmaint = 500\,\kms$ for various \NCIV and \vout. The left panel shows the \CIV spectra for $\NCIV = 10^{13}$--$10^{16}\,\unitNHI$ at a fixed outflow velocity of $\vout = 500\,\kms$, while the right panel shows the variation of the \CIV spectra for $\vout = 100$, 300, 500, and $1000\,\kms$ at $\NCIV = 10^{16}\,\rm cm^{-2}$. Overall, the dependence of the \CIV spectrum on \NCIV and \vout is similar to that in Figure~\ref{fig:Fig_8}. Both transitions exhibit blueshifted absorption features over the velocity range from $-\vout$ to $0\,\kms$, and the P-Cygni profile becomes increasingly pronounced with increasing \NCIV. 
Interestingly, the initially mixed K and H components can be partially separated and become distinguishable through scattering in an outflowing medium.
When $\vout = 1000\,\kms$, the \CIV profile is dominated by a strong H component, which can be interpreted as an asymmetric, redshifted single emission line.

\subsubsection{C IV Doublet Ratio}
\label{sec:doublet_ratio}

\begin{figure*}
    \centering
    \includegraphics[width=\linewidth]{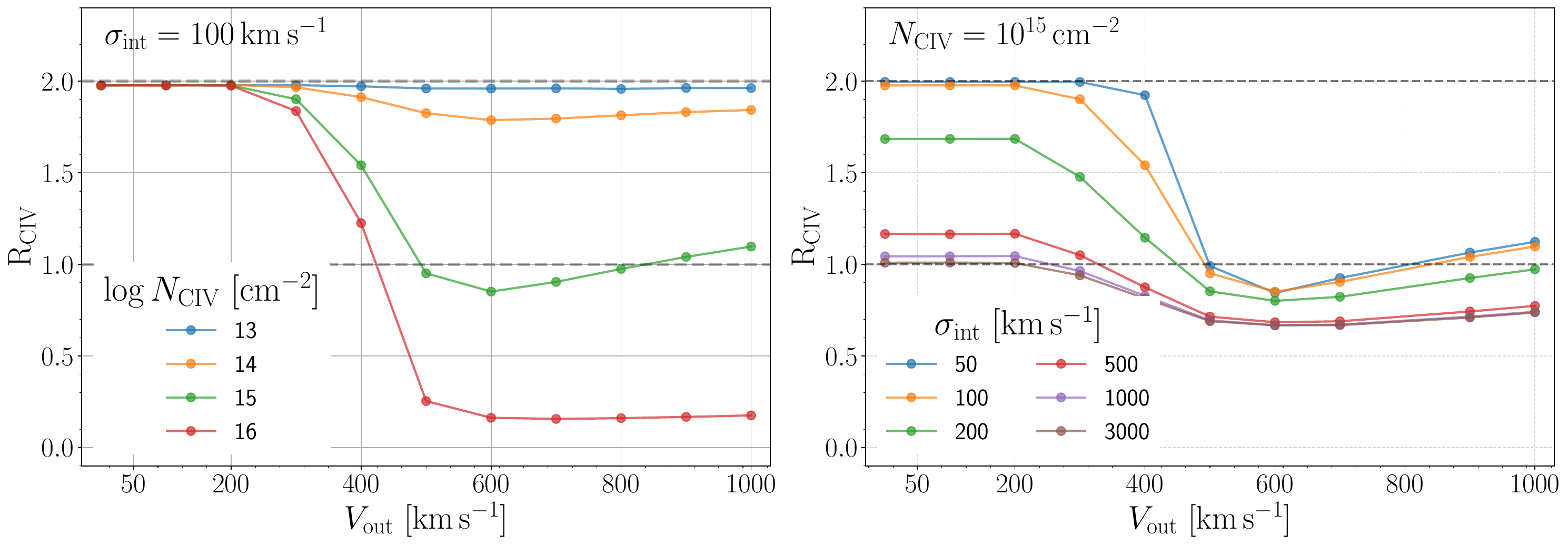}
    \caption{\CIV doublet ratio \RCIV as a function of outflow velocity \vout for various \CIV column densities \NCIV and intrinsic emission widths \sigmaint. The velocity dispersion is fixed at \sigmaR = 11.8\,\kms. The left panel shows \RCIV at $\sigmaint = 100\,\kms$ to highlight the effect of outflows. The colored solid lines correspond to $\NCIV = 10^{13}$–$10^{16}\,\unitNHI$. The right panel shows \RCIV at $\NCIV = 10^{15}\,\unitNHI$, where the ratio is sensitive to outflows. The colored solid lines correspond to $\sigmaint = 50$–$3000\,\kms$. In both panels, the dashed lines mark the intrinsic doublet ratio (\RCIV = 2) and the limiting value determined by the intrinsic emission width (\RCIV = 1).}
    \label{fig:Fig_10}
\end{figure*}

We assume an intrinsic doublet \CIV ratio of 2, defined as the ratio of the K and H line fluxes, based on their atomic transition probabilities (see \S~\ref{sec:cross_section}). However, as discussed in the previous section, a fast ($\geq 500\,\kms$) outflow can mix K- and H-line photons even when the two components are intrinsically well resolved (see Figure~\ref{fig:Fig_8}). In this regime, the emergent spectrum becomes dominated by the H line, reducing the doublet ratio below its intrinsic value. This behavior suggests that the \CIV doublet ratio can serve as a useful indicator of fast outflows. To quantify the doublet ratio and assess its effectiveness as an outflow indicator, we define \RCIV as follows:
\begin{equation}
    R_{\ion{C}{IV}} = \frac{\int^{\lambda_{0,\rm K} + \lambda_c}_{\lambda_{0,\rm K} - \lambda_c} F(\lambda)d\lambda}{\int_{\lambda_{0,\rm H} - \lambda_c}^{\lambda_{0,\rm H} + \lambda_c} F(\lambda)d\lambda},
    \label{eqn_gaussian_analytic_a}
\end{equation}
where $F(\lambda)$ is the flux as a function of wavelength $\lambda$, and $\lambda_c = (\lambda_{0,\rm H} - \lambda_{0,\rm K})/2$ is the criterion for the K and H line regions, corresponding to $\Delta V = 250\,\kms$ in Section~\ref{sec:Gaussian}.
In Figure~\ref{fig:Fig_10}, we show \RCIV as a function of \vout for various \sigmaint and \NCIV.

The left panel of Figure~\ref{fig:Fig_10} illustrates the dependence of \RCIV on \vout for different \CIV column densities at a fixed intrinsic line width of $\sigma_{\rm int}=100\,\kms$. At the lowest column density, $\NCIV=10^{13}\,\unitNHI$, \RCIV remains close to its intrinsic value of 2 over the entire range of \vout, indicating that resonance scattering is negligible. At higher column densities ($\NCIV\geq10^{14}\,\unitNHI$), however, the doublet ratio becomes increasingly sensitive to \vout. For $\vout\lesssim200\,\kms$, \RCIV remains close to 2 even at the highest column density, $\NCIV=10^{16}\,\unitNHI$, because the outflow velocity is insufficient to efficiently mix K- and H-line photons through resonance scattering. As \vout approaches or exceeds the doublet velocity separation, \RCIV decreases sharply, with the decline becoming more pronounced at higher column densities. At $\vout>600\,\kms$, however, \RCIV becomes nearly constant, particularly at the highest column density. This behavior occurs because most K-line photons undergo multiple scatterings at high column densities. Interestingly, at the intermediate column density of $\NCIV=10^{15}\,\unitNHI$, \RCIV increases slightly with increasing \vout beyond $600\,\kms$, as higher outflow velocities reduce the effective optical depth\footnote{A similar trend is also seen in the Mg~II RT simulation by \citet{Chang24}. They explained that as \vout increases, the optical depth decreases, which reduces the probability of scattering (see Equations~(14) and~(20) in \citealt{Chang24}).}. This suggests that the \CIV doublet ratio, \RCIV, has the potential to serve as a tracer of strong outflows.

In the right panel of Figure~\ref{fig:Fig_10}, we show \RCIV for intrinsic line widths of $\sigmaint = 50$--$3000\,\kms$ at a fixed column density of $\NCIV = 10^{15}\,\unitNHI$. When $\vout \leq 200\,\kms$, the outflow velocity is insufficient to efficiently mix the K- and H-line photons through resonance scattering, and \RCIV depends primarily on \sigmaint, approaching unity as \sigmaint increases (see details in Appendix~\ref{Appendix:Intrinsic}). 
Accordingly, narrow intrinsic profiles retain \RCIV close to its intrinsic value of 2, whereas broader profiles exhibit lower doublet ratios even at small \vout because the K and H components are more strongly blended.
As \vout increases to approximately $400$--$600\,\kms$, \RCIV declines rapidly and eventually falls below unity. In most cases, \RCIV reaches a minimum near $\vout \sim 600\,\kms$ and then gradually increases at higher outflow velocities.
These results indicate that a doublet ratio below unity is a useful signature of strong outflows.
When $\sigmaint \lesssim 500\,\kms$, the doublet ratio is sensitive to \vout and can serve as a tracer of outflows with $\vout\gtrsim 200\,\kms$ in systems with relatively low-\sigmaint, such as star-forming galaxies. In contrast, when $\sigmaint \gtrsim 500\,\kms$, \RCIV remains nearly constant at $\vout > 600\,\kms$, making it difficult to constrain the outflow velocity in systems with high-\sigmaint, such as AGNs. Nevertheless, $\RCIV < 1$ remains a useful indicator of strong outflows (i.e., $\vout \gtrsim 500\,\kms$).

We have demonstrated above that the \CIV doublet ratio serves as a diagnostic of fast outflows in the optically thick regime (i.e., $\NCIV \geq 10^{14}\,\unitNHI$). 
In particular, this interpretation is most robust in low-\sigmaint systems, where the K and H components are clearly separated, allowing a direct assessment of scattering in an outflowing medium. 
This may help to explain the range of \CIV doublet ratios observed in star-forming galaxies, which deviate from the intrinsic value of 2 \citep[e.g.,][]{Stark2015,Vanzella2016,senchyna2017,Berg2019B,Izotov2024,Tang2024,Topping2025}.
In Section~\ref{sec:trace_outflow}, we will further discuss the use of \RCIV as a tracer of fast, warm outflows at $\sigmaint = 100\,\kms$.

\begin{figure*}
    \centering
    \includegraphics[width=\linewidth]{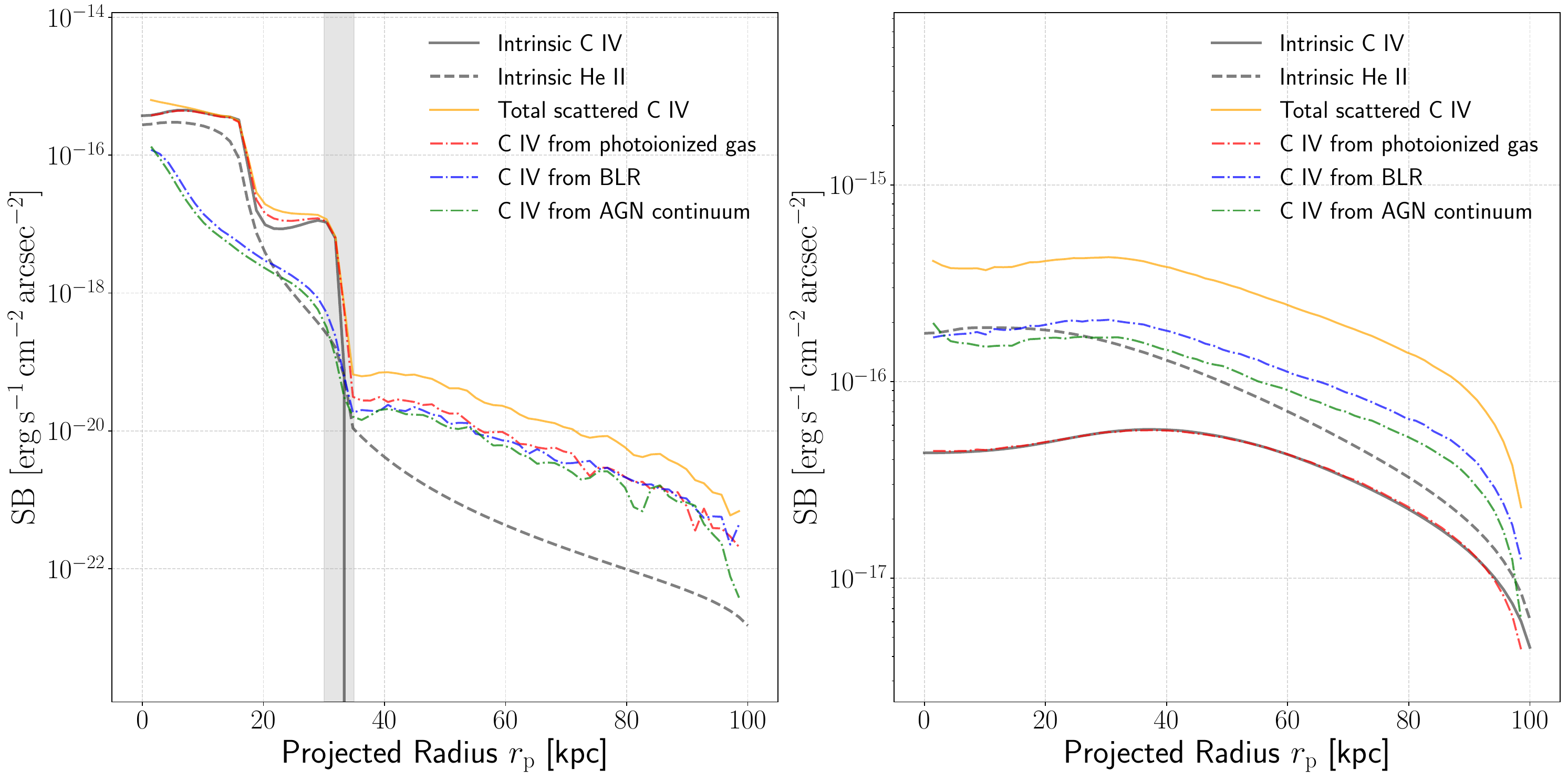}
    \caption{Radial surface brightness (SB) profiles of \CIV and \HeII emission. The left and right panels show the ionization-bounded and density-bounded cases, with ionizing luminosities at 1 Ryd of $\nu L_{\nu}=10^{44}$ and $10^{46}\,{\rm erg\,s^{-1}}$, respectively. In the left panel, the gray region marks the hydrogen ionization front. In both panels, the intrinsic SB profiles of \CIV and \HeII are shown by black solid and dashed lines, respectively. The colored dash-dotted lines indicate the contributions from the three input photon sources after resonance scattering in an outflow with $\vout = 500$\,\kms: the in-situ intrinsic \CIV emission produced within the photoionized gas (red), broad \CIV line emission from the AGN (blue), and continuum radiation near the \CIV line (green). The orange solid lines represent the simulated \CIV SB profiles obtained by combining all three photon sources. The velocity dispersion and the intrinsic emission-line width of the central source are fixed at $\sigmaR = 30$\,\kms and $\sigmaint = 3000$\,\kms, respectively.
    }    
    \label{fig:Fig_11}
\end{figure*}

\subsection{Spatially extended C IV halos around quasars}
\label{sec:Spatial_distribution}

Spatially extended nebulae of \CIV$\lambda\lambda$1548, 1551 and \HeII$\lambda$1640 have been observed around quasars at $z > 2$ \citep{Guo2020,Travascio2020,Fossati2021,Sabhlok2024,Galbiati2026}. 
Because \CIV and \HeII have similar ionization energies, 47.9 and 54.4\,eV, respectively, both lines are expected to trace warm, highly ionized gas in the CGM at $T \sim 10^5\,{\rm K}$. 
However, their radiative properties are fundamentally different. 
\HeII$\lambda$1640, the Balmer-$\alpha$ transition of He~II, is a non-resonance line that traces the local recombination emissivity of He~III, 
whereas the \CIV resonance doublet is predominantly produced by collisional excitation and can additionally undergo resonance scattering by \CIV atom. 
As a result, the spatial distribution of \CIV emission does not necessarily need to follow that of \HeII. \CIV photons can be redistributed to larger radii by scattering, while \HeII photons escape without resonance scattering. 
This distinction is important because photoionization models alone often struggle to reproduce the observed \CIV and \HeII luminosities and their relative spatial distributions unless crude approximations of resonance scattering are taken into account. 
For example, \citet{Fossati2021} detected extended \CIV emission out to $\simeq50$\,kpc in a stack of 27 bright quasars at $z=3$--$4.5$, with a tentative detection of \HeII out to $\simeq40$\,kpc, but found that the observed spatial extents and line ratios are difficult to reproduce with pure photoionization models and require the addition of photon pumping. 
A similar result was also shown by \citet{Obreja2024} when post-processing a cosmological simulation of a massive halo expected to host a $z\sim3$ quasar. Only including the approximated treatment of broad-line region photon scattering allowed the authors to simultaneously match the observed levels of extended \HeII and \CIV emission around quasars.
In this section, we therefore first compute the intrinsic \CIV and \HeII surface-brightness profiles expected from photoionization\footnote{
To avoid double-counting scattering effects, we turn off the line-pumping treatment in \texttt{CLOUDY}, which approximates resonance-line scattering as a single-scattering process.}, and then post-process the \CIV emission with radiative transfer simulations to quantify how resonance scattering modifies its spatial distribution.

\subsubsection{Photoionization Modeling}

To characterize the intrinsic spatial distributions of \CIV and \HeII emission, we compute their emissivities using the public photoionization code \texttt{CLOUDY} (v17.03; \citealt{Ferland2017}).
We assume an exponentially decreasing hydrogen density profile, $n_{\rm H}(r)$, spanning 1--100\,kpc, and fix the total hydrogen column density to $N_{\rm H}=10^{22}\,\unitNHI$.
For the central ionizing source, we adopt the QSO spectral energy distribution (SED) used in \citet{Fab2015b}, which is a combination of power-laws obtained from quasar template spectra \citep{Vanden2001,Strateva2005,Richards2006,Lusso2015}.
The SED is normalized to ionizing luminosities of $\nu L_\nu = 10^{44}$ and $10^{46}\,{\rm erg\,s^{-1}}$ at 1\,Rydberg for the ionization-bounded and density-bounded cases, respectively.
The adopted column density is chosen to reproduce the observed range of \HeII$\lambda1640$ luminosities, $L_{\rm He\,II}=10^{42}$--$10^{43}\,{\rm erg\,s^{-1}}$ \citep[e.g.,][]{Zhang2023,Galbiati2026}.
Further details of the photoionization setup are provided in Appendix~\ref{Appendix:Photoionization}.

Figure~\ref{fig:Fig_11} shows the radial surface-brightness profiles of \CIV and \HeII for the ionization-bounded and density-bounded cases.
The intrinsic profiles predicted by photoionization alone depend strongly on the ionization state of the nebula.
In the ionization-bounded case, \CIV emission is intrinsically brighter than \HeII inside the ionization front, corresponding to projected radii $r_{\rm p} \lesssim 30\,\mathrm{kpc}$. However, its surface brightness declines steeply near the ionization front, making the intrinsic \CIV profile more compact than that of \HeII.
In the density-bounded case, \HeII remains much brighter than \CIV at all radii.
Thus, in both photoionization-only models, the intrinsic \CIV profile is either more compact than the \HeII profile or substantially fainter than \HeII. 
In other words, the photoionization-only models considered above do not naturally reproduce the observation that \CIV emission is more spatially extended than \HeII, and is often detected without any \HeII counterpart.
To address this, we will therefore show the \CIV surface brightness considering its resonance scattering in the following section.

\subsubsection{Resonantly scattered C~IV halo}
\label{sec:resonantly_scattered_civ}

We combine the \texttt{CLOUDY} photoionization models with RT-scat to compute the \CIV surface-brightness profile, including the effects of resonance scattering.
The scattering geometry is composed of a spherical \CIV halo, whose radial \CIV distribution is derived from the \texttt{CLOUDY} calculations.
The halo is illuminated by three sources of \CIV photons: the in-situ intrinsic \CIV emission produced within the photoionized gas, broad \CIV emission from the broad-line region, and the AGN continuum near the \CIV line center.
The luminosities of these three components are listed in Table~\ref{tab:Luminosity_table}.
For the RT-scat calculation, we adopt a velocity dispersion of $\sigmaR = 30\,\kms$, assuming the \CIV-bearing gas has turbulent motions similar to those assumed for \Lya emission around AGNs \citep{prochaska2013,Fab2015b}, and a representative maximum outflow velocity of $V_{\rm out}=500\,\kms$.
We note that the scattered \CIV surface-brightness profiles are plotted over $0<r_{\rm p}<100\,{\rm kpc}$, since the surface brightness rises steeply toward $r_{\rm p}=0$ and drops below the lower limit of the vertical axis at $r_{\rm p}=100\,{\rm kpc}$.
Further details of the combined \texttt{CLOUDY} and RT-scat setup are provided in Appendix~\ref{Appendix:Photoionization}.

Figure~\ref{fig:Fig_11} shows the \CIV surface-brightness profiles after post-processing with resonance scattering. In both the ionization-bounded and density-bounded cases, the scattered \CIV profile is brighter and more extended than the intrinsic photoionization-only profile, and its surface brightness exceeds that of \HeII over most of the radial range.
In the ionization-bounded case, the scattering effect is marginal within the ionization front, $r_{\rm p}\lesssim 30\,{\rm kpc}$, where the scattered profile remains comparable to the photoionization-only model.
Beyond the ionization front, however, the scattering effect becomes conspicuous, as \CIV photons are redistributed to larger radii through scattering in the outflowing gas. The resulting profile is more extended than that of the photoionization-only model and remains brighter than \HeII even at large radii. 
In the density-bounded case, the scattering effect is prominent at all radii, where the scattered \CIV surface-brightness profile lies above the \HeII profile. 
Given the fundamental differences in the radiative properties of \CIV and \HeII, resonance scattering provides a viable mechanism for the observed extended \CIV halos, which cannot be accounted for by photoionization alone.

The relative contributions of the three \CIV photon sources depend on the ionization state of the halo. 
In the ionization-bounded case, the \CIV surface brightness within the ionization front, $r_{\rm p}\lesssim 30\,{\rm kpc}$, is dominated by photons produced locally in the photoionized gas. 
At larger radii, however, scattered photons from the photoionized gas, broad-line region, and AGN continuum contribute at comparable levels.
In the density-bounded case, the contribution from photoionized gas is weak, and the scattered broad-line and continuum photons from the AGN dominate the emergent \CIV halo.
These results show that resonance scattering of AGN radiation, together with locally produced \CIV photons, can play an important role in forming extended \CIV nebulae around quasars.

In summary, our results demonstrate that resonance scattering can play a central role in shaping the spatial extent of \CIV emission around AGNs.
While photoionization alone does not naturally produce a \CIV halo that is more extended than \HeII, the inclusion of resonance scattering generates extended \CIV surface-brightness profiles over several tens of kpc, comparable to the observed \CIV halos around quasars.
This occurs because \CIV photons produced in the photoionized gas, together with broad emission and continuum photons from the AGN, can be redistributed to larger radii by resonance scattering.
Since the efficiency of this redistribution depends on the outflow velocity of the \CIV gas, we discuss the velocity dependence of the \CIV surface-brightness profile in Section~\ref{sec:SB_outflow}.

\begin{table}
    \centering
    \renewcommand{\arraystretch}{1.5}
    \caption{Luminosities of \HeII$\lambda$1640 and three \CIV emission components (see details in Appendix~\ref{Appendix:Photoionization}).}
    \begin{tabular}{c c c}
        \hline
        \hline
        Luminosity~$\rm [erg \, s^{-1}]$  & Ionization-bounded$^{*}$ & Density-bounded$^{\dagger}$ \\
        \hline
        $L_{\rm He\,II}$ & $1.00 \times 10^{42}$ & $1.05 \times 10^{43}$ \\
        \hline
        $L_{\rm CIV, gas}$ & $1.90 \times 10^{42}$ & $5.17 \times 10^{42}$ \\
        \hline
        $L_{\rm CIV, BLR} $ & $1.66 \times 10^{42}$ & $1.64 \times 10^{44}$ \\
        \hline
        $L_{\rm CIV, cont}$ & $5.90 \times 10^{42}$ & $5.90 \times 10^{44}$ \\
        \hline
    \end{tabular}
    \\
    \footnotesize
    QSO SED is normalized to $\nu L_{\nu} = 10^{44.0}\,{\rm erg\,s^{-1}}$ ($*$) and $10^{46.0}\,{\rm erg\,s^{-1}}$ ($\dagger$) at 1\,Ryd.
    \label{tab:Luminosity_table}
\end{table}

\section{Discussion}\label{sec:Discussion}

\subsection{C IV doublet ratio as a tracer of fast outflow}\label{sec:trace_outflow}

In \S~\ref{sec:Gaussian}, we show that the doublet ratio exhibits significant variation when the outflow velocity is faster than the velocity separation of the doublet \citep[see also][]{Yoo2002,Chang24}.
Even though the \CIV doublet ratio \RCIV depends on the intrinsic width, a ratio less than unity is clear evidence of strong outflow as shown in Figure~\ref{fig:Fig_10}. 
This small doublet ratio has been observed in various stellar objects \citep{Feibelman1983,Michalitsianos1988,Michalitsianos1992,Cauley2016}, 
as well as in the strong, saturated P-Cygni profiles in OB stars and Wolf-Rayet stars \citep{Willis1986,Crowther2022}, where a strong outflow at $T \sim 10^5\, \rm K$ is able to exist.
Furthermore, recent \CIV observations of star-forming galaxies show a wide range of \RCIV values from 0.3 to 2.4 \citep[e.g.,][]{Stark2015,Vanzella2016,senchyna2017,Berg2019A,Berg2019B,Matthee2022,Tang2024,Izotov2024,Witstok2021,Witstok2025}.
Moreover, \cite{Berg2019B} showed that the \CIV profile is broader than those of He~II and \OIII emission lines, which are non-resonance lines with similar ionization energies.
Thus, \RCIV can serve as a tracer of warm gas outflows in star-forming galaxies through radiative transfer analysis.

\begin{figure}
    \centering
    \includegraphics[width=\linewidth]{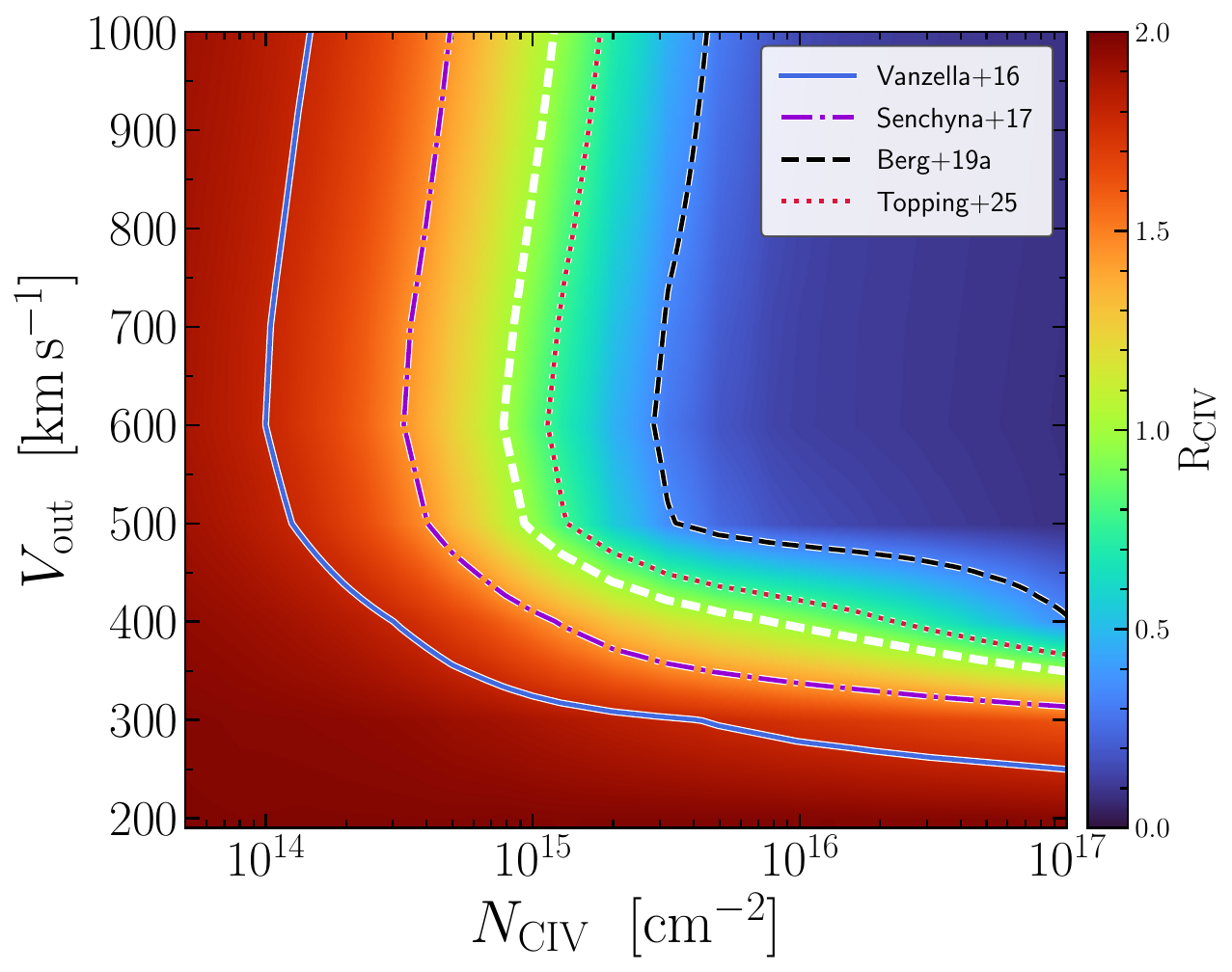}
    \caption{Two-dimensional map of the \CIV doublet ratio \RCIV as a function of \CIV column density $\NCIV$ and outflow velocity $\vout$, with velocity dispersion and intrinsic source width fixed at $\sigmaR = 30$\,\kms and $\sigmaint = 100$\,\kms, respectively. Contours indicate observed \RCIV values from star-forming galaxies: 1.79 (blue solid, \citealt{Vanzella2016}), 1.45 (magenta dash-dotted, \citealt{senchyna2017}), 0.78 (red dotted, \citealt{Topping2025}), and 0.35 (black dashed, \citealt{Berg2019A}). The white dashed contour marks \RCIV = 1. }
    \label{fig:Fig_12}
\end{figure}

Figure~\ref{fig:Fig_12} presents \RCIV as a 2D map in the \NCIV--\vout parameter space, with observed \RCIV values from star-forming galaxies overlaid to identify the parameter combinations that reproduce the observations. At column densities $\NCIV = 10^{14}$--$10^{15}\,\unitNHI$, relatively strong outflows ($\vout \gtrsim 300\,\kms$) are required to reproduce the observed values of $\RCIV = 1.79$ for ID11 at $z = 3.12$ \citep{Vanzella2016} and $\RCIV = 1.45$ for SB\,82 \citep{senchyna2017} in the local Universe.
However, even with the strongest outflows in our models ($\vout = 1000\,\kms$), \RCIV does not fall below unity in this \NCIV regime, as indicated by the white contour in Figure~\ref{fig:Fig_12}.
In contrast, at higher column densities ($\NCIV = 10^{15}$--$10^{16}\,\unitNHI$), scattering becomes significant, and \RCIV can drop below unity. This regime reproduces the observed values of $\RCIV = 0.78$ for A1703-zd6 at $z = 7.04$ \citep{Topping2025} and $\RCIV = 0.35$ for the local galaxy J094530 \citep{Berg2019A}, requiring outflow velocities of $\vout \gtrsim 400\,\kms$.
These results demonstrate that \RCIV is sensitive to both column density and outflow velocity, and can therefore serve as a tracer of the kinematics and physical properties of outflowing gas.

Importantly, the diagnostic power of the \CIV doublet ratio is significantly enhanced when combined with tracers of other gas phases.
While \CIV probes warm gas ($T \sim 10^5\,\rm K$), other resonance transitions are sensitive to gas at different temperatures.
Low-ionization lines such as Mg~II and \Lya trace the cool gas phase ($T \lesssim 10^4\,{\rm K}$), whereas high-ionization lines such as N~V and O~VI probe hotter gas ($T \gtrsim 10^{5.5}$--$10^6\,{\rm K}$).
In addition, the observed doublet ratio, particularly when interpreted in conjunction with the velocity separation of the doublet components, can provide insight into the gas phase and its kinematic structure \citep{Chang24}. Given that gas within and around galaxies is intrinsically multiphase, comprising both cool and hot components \citep{Veilleux2005,Veilleux2020,Faucher2023,GronkeSchneider2026}, the simultaneous analysis of multiple resonance lines would provide a powerful means of constraining the physical conditions and kinematics of the multiphase gas.

\subsection{C IV surface brightness \& outflow}
\label{sec:SB_outflow}

\begin{figure*}
    \centering
    \includegraphics[width=\linewidth]{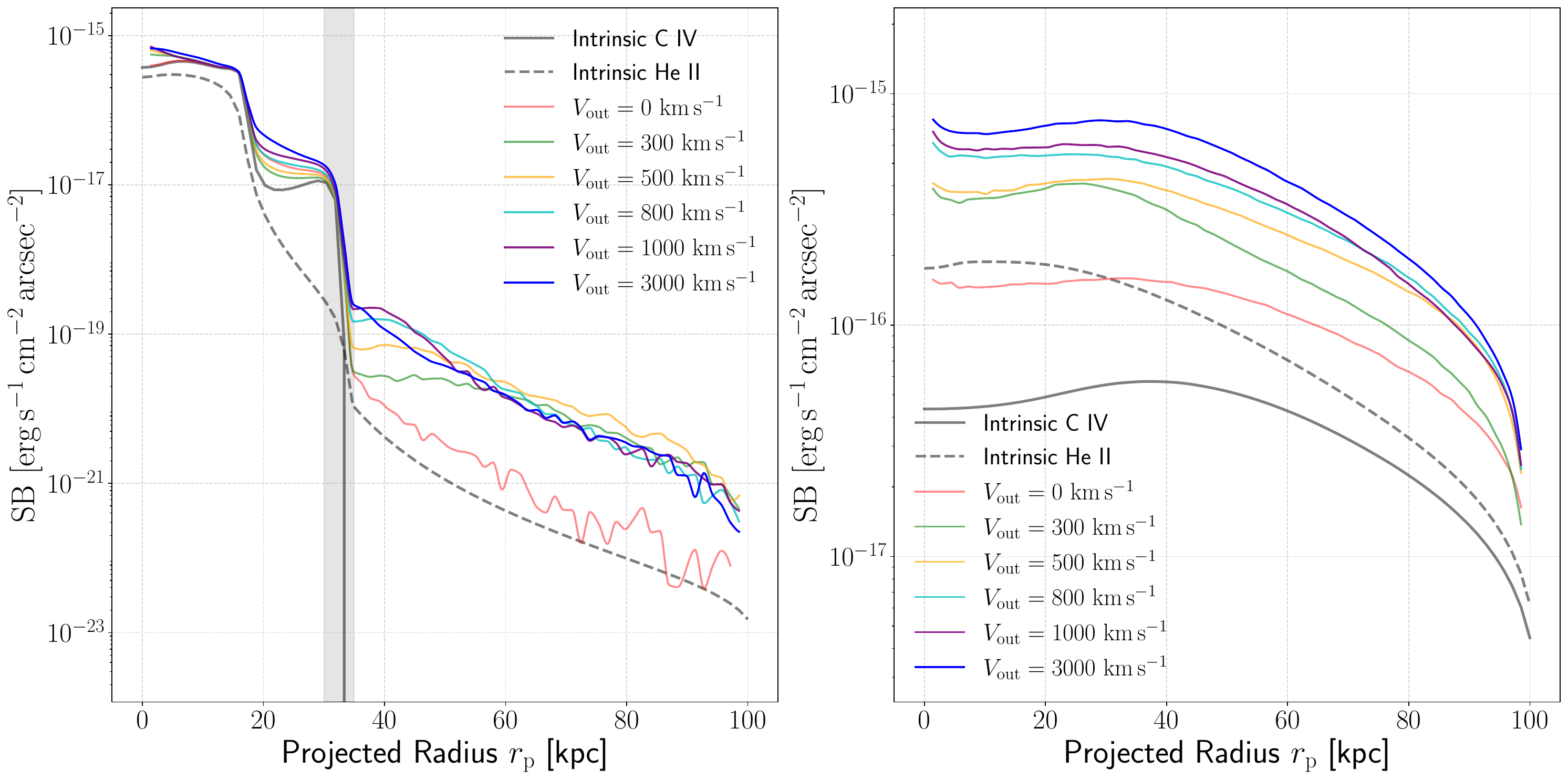}
    \caption{Radial surface brightness (SB) profiles of \CIV and \HeII emission for various outflow velocities \vout, following the same convention as in Figure~\ref{fig:Fig_11}. 
    The SB profiles of the total scattered \CIV are shown for \vout = 0 (red), 300 (green), 500 (yellow), 800 (cyan), 1000 (purple), and 3000 (blue)\,\kms. 
    The yellow solid line is identical to the total surface-brightness profile in Figure~\ref{fig:Fig_11}, which is represented by the orange solid line.}
    \label{fig:Fig_13}
\end{figure*}

Resonance scattering plays a crucial role in producing the extended \CIV surface brightness around AGNs compared to the non-resonant \HeII$\lambda1640$ emission, as shown in Section~\ref{sec:Spatial_distribution}. The efficiency of scattering is governed by both the kinematics and the physical properties of the scattering medium. 
In particular, an outflowing, optically thick medium can expose photons to a broad range of line-of-sight velocities when the radial velocity is non-uniform. 
This effect also allows intrinsically broad emission of resonance lines, such as Ly$\alpha$ and Mg~II, to scatter over a wider frequency range \citep[e.g.,][]{Zheng2002,Verhamme2006,Chang24}.
This is a similar behavior to \CIV as shown in Figures~\ref{fig:Fig_8} and \ref{fig:Fig_9}.
The scattered photons in the broad frequency range contribute to the spatial expansion of \Lya nebulae \citep{Chang23}.
Observationally, AGN-driven outflows have also been observed in the CGM in association with extended \CIV and \HeII nebulae \citep[e.g.,][]{Sabhlok2024}, and \citet{Rueda_Vargas2026} found strong correlations between the ISM-scale outflow velocity and both the size and luminosity of the surrounding \Lya nebulae in five QSOs at $z\sim2$--$3$, arguing that similar trends might be expected on CGM scales.
Motivated by these results, we explore in this section how the outflow velocity affects the \CIV surface-brightness profiles, using models in Section~\ref{sec:Spatial_distribution}.

Figure~\ref{fig:Fig_13} shows the \CIV surface-brightness profiles of the ionization-bounded and density-bounded cases for outflow velocities $\vout = 0-3000\,\kms$. The resonantly scattered \CIV emission in the outflowing medium is brighter than that in the static case ($\vout = 0\,\kms$) out to the outskirts.
However, the efficiency of scattering shows different behavior depending on the ionization condition.
In the density-bounded case, the surface-brightness profile of \CIV emission brightens with increasing \vout at most radii.
The ionization-bounded case presents similar behavior within the ionization front ($r_{\rm p} \lesssim 30\,$kpc).
At larger radii ($> 60\,$kpc), the profiles in outflowing cases are similar and clearly more extended than those in the static case.
These results indicate that resonance scattering in outflow enhances the \CIV surface brightness and contributes significantly to spatial expansion.

Our results are also directly relevant to observations of extended \CIV and \HeII haloes around AGNs. 
At current surface brightness limits, \HeII nebulae are generally more compact and fainter than their \CIV counterparts \citep{Guo2020,Zhang2023,Sabhlok2024,Galbiati2026}. In some cases, extended \CIV emission is detected with no \HeII counterpart \citep{Travascio2020}.
Resonance scattering can account for this difference by redistributing C~IV photons to larger radii, whereas He~II directly traces the local emissivity.
In Figure~\ref{fig:Fig_13}, the \CIV profiles at $\vout \geq  300 \kms$ are much brighter than He~II in the outer radius $> 40\,$kpc, while the static case profile is comparable to the He~II profile.
These results show that resonance scattering in outflowing gas can be important for interpreting highly ionized nebular emission.

More generally, the spatial expansion of resonance lines is determined by both the distribution and kinematics of the scattering gas.
Velocity gradients systematically shift resonance photons in frequency and reduce trapping \citep{Bonilha1979}. Observationally, the similar distributions of [C~II] and Ly$\alpha$ haloes, along with the larger Ly$\alpha$ velocity offsets associated with extended [C~II] emission, connect the kinematics of cold gas to the spatial redistribution of Ly$\alpha$ photons \citep{Fujimoto2020}. Our results extend this picture to warm ionized gas. 
At fixed gas distribution, outflow velocity extends the C~IV scattering region. In real systems, however, outflows also change the amount and ionization state of the scattering gas.
\citet{Costa2022} found that AGN feedback initially facilitates extended Ly$\alpha$ emission by clearing dusty, H~I gas from the nucleus, whereas stronger or sustained feedback lowers the H~I column density in the halo and makes the Ly$\alpha$ emission more compact.
C~IV may therefore provide a complementary probe of warm outflows when Ly$\alpha$ scattering becomes less effective.

\subsection{Implications for C~IV emission in Little Red Dots}\label{sec:LRD}

Recent James Webb Space Telescope (\textit{JWST}) observations have revealed a population of compact sources in the early Universe, Little Red Dots (LRDs) with broad, asymmetric Balmer lines and V-shaped spectral energy distributions with blue UV and
red optical continuum \citep{Harikane2023,Ulber2023,Kocevski2023,Maiolino2024,Matthee2024,
Greene2024,Kocevski2025}.
Their compactness and broad emission lines suggest the presence of AGN, but many LRDs lack traditional AGN signatures, such as strong X-ray and radio emission or a prominent infrared excess \citep{Sacchi2025,Gloudemans2025,Xiao2025}.
This has motivated the idea that the radiative transfer effects of Balmer lines, such as Thomson and resonance scattering, produce broad, asymmetric Balmer lines without AGN properties \citep{Rusakov2026,chang2026,Sneppen2026}. In other words, Thomson-thick gas (i.e., electron column density $N_{\rm e}> 10^{24} \unitNHI$) is required to generate broad wings around Balmer lines.

C~IV emission has now been detected in several LRDs \citep{Labbe2024,Akins2025,Ando2026,Ji2026,Tang2026}.
Strong C~IV accompanied by weak or undetected He~II has often been interpreted as evidence for photoionization by low-metallicity massive stars rather than by a typical AGN spectrum. However, current observations do not uniquely determine whether C~IV is produced by the compact LRD component or its host galaxy. Moreover, the ionizing source, the intrinsic emission region, and the gas through which the photons propagate are physically different.

We show that the relative spatial distributions of C~IV and He~II can provide additional constraints in Section~\ref{sec:SB_outflow}. 
While non-resonant He~II approximately traces the emission region, 
C~IV photons can be spatially extended by scattering with warm gas. 
C~IV emission that is more extended than He~II may therefore indicate scattering rather than a difference in the intrinsic emission regions alone. 
Their line profiles provide a complementary diagnostic.
If C~IV and the Balmer lines propagate through the Thomson-thick gas, they may initially have similar broad wings, and the doublet is mixed by Thomson scattering, whereas subsequent C~IV resonance scattering can additionally produce velocity offsets, asymmetry, absorption features, and changes in the doublet ratio. In particular, $R_{\rm CIV}<1$ would indicate an optically thick, fast warm outflow as shown in Section~\ref{sec:trace_outflow}.

Future observations that resolve both the spatial and spectral distribution of C~IV will therefore be important for identifying its origin in LRDs. Joint measurements of C~IV, He~II, and the Balmer lines can separate the effects of the ionizing spectrum, electron scattering, and C~IV resonance transfer, providing a direct probe of the multiphase gas surrounding LRDs.

\section{Conclusions}\label{sec:Conclusion}

This paper extended the 3D Monte Carlo radiative transfer code \texttt{RT-scat} \citep{Chang24} to include the \CIV resonance doublet at $\lambda\lambda 1548,1551$. Using a simple spherical geometry, we investigated the effects of \CIV resonance scattering on the line formation, the variation of the doublet ratio, and the spatial distribution of \CIV emission.

Our main conclusions are summarized as follows:
\begin{itemize}
    \item \textbf{Resonance scattering shapes the \CIV spectrum.} As the line-center optical depth \tauo increases, the profile evolves from a single Gaussian to a double-peaked shape with central suppression, broadening as $\rm{FWHM}/\sigmaR \approx 1.092 + 1.061\sqrt{\ln\left[1 + \left(\tauo/1.505\right)^{2.788}\right]}$. 
    In an outflowing medium the doublet develops P-Cygni features, and once \vout exceeds the K--H separation ($\vsep \approx 500\,\kms$), K-line photons are redistributed around the H component---naturally explaining the observed dominance of the \CIV$\lambda1551$ component in star-forming galaxies.

    \item \textbf{The doublet ratio \RCIV traces fast, warm outflows.}
    \RCIV decreases from its intrinsic value of $\sim 2$ as \NCIV and \vout increase, falling below unity for strong outflows in the optically thick medium (see \S~\ref{sec:doublet_ratio} and Figure~\ref{fig:Fig_12}). 
    Since intrinsic blending alone cannot drive \RCIV below unity, $\RCIV < 1$ can arise from resonance scattering in strong outflows. Our \NCIV--\vout map reproduces the range of \RCIV observed in star-forming galaxies, demonstrating its diagnostic power on real data.

    \item \textbf{Scattering produces extended \CIV halos.} 
    Photoionization alone cannot make \CIV more extended than \HeII$\lambda1640$: the intrinsic \CIV profile is either more compact or much fainter than \HeII. 
    Resonance scattering redistributes \CIV photons---from in-situ photoionized gas, the broad-line region, and AGN continuum---to larger radii, producing extended \CIV halos over several tens of kpc that exceed \HeII and brighten with \vout, comparable to the \CIV halos observed around quasars (see Section~\ref{sec:Spatial_distribution} and Figure~\ref{fig:Fig_13}).
\end{itemize}

Several open questions remain for future work. Other UV metal resonance doublets, such as O\,VI $\lambda\lambda$1032,\,1038 and N\,V $\lambda\lambda$1239,\,1243, are expected to show similar behavior and, combined with \CIV, can constrain the multiphase structure of the warm CGM at $T \geq 10^{5}\,$K. More realistic geometries---including gas clumpiness, anisotropic optical depths, and polarization effects---may also introduce more complex behavior in the emergent doublet ratios \citep{Seon2024,Chang24}. Integrated models that combine multiple resonance lines with different ionization energies will further help break degeneracies of modeling and trace distinct gas phases of the outflow \citep{Li2026}.
Beyond these methodological extensions, alternative powering mechanisms such as shock heating, which can collisionally ionize and heat the CGM to $T \sim 10^5$--$10^6\,$K and produce \CIV and \HeII emission without an external photoionizing field \citep{Allen2008, Cabot2016}, deserve exploration. Fully reproducing the observed spatial extent and line ratios of \CIV and \HeII around quasars may require both resonance scattering and shock heating.

\section*{Acknowledgements}

SJC acknowledges support from the ERC synergy grant 101166930 - RECAP. 
J. Lim and K.-I. Seon are supported by the Korea Astronomy and Space Science Institute (KASI) through the R\&D program (Project No.~2026-1-861-02) supervised by the Ministry of Science and ICT.
MG thanks the European Union for support through ERC-2024-STG 101165038 (ReMMU).
H.-W. Lee gratefully acknowledges the support of the National Research Foundation of Korea (NRF) grant funded by the Korean government (No. NRF-2023R1A2C1006984).

\section*{Data Availability}

Data related to this work will be shared on reasonable request to the corresponding author.

\bibliographystyle{mnras}
\bibliography{reference}

\appendix
\section{Analytic solution for C~IV doublet ratio depending on intrinsic emission width}\label{Appendix:Intrinsic}

\begin{figure}
    \centering
    \includegraphics[width=\linewidth]{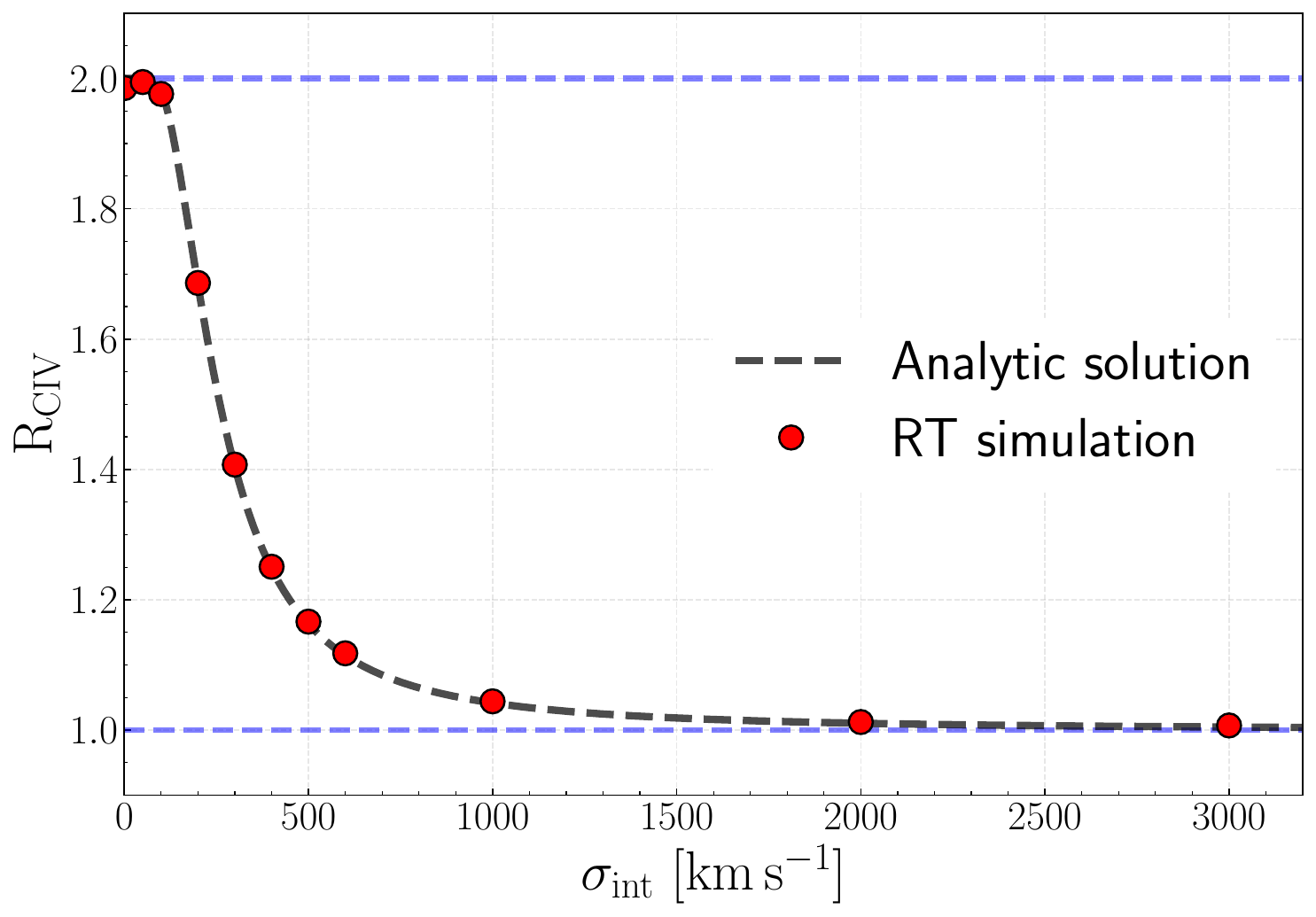}
    \caption{\CIV doublet ratio \RCIV as a function of the intrinsic emission width for \sigmaint = $0\text{--}3000~\kms$. The black dashed line represents the analytic solution as described in Equation~\ref{eqn_gaussian_analytic_b}, and the blue dashed horizontal lines indicate the limits of \RCIV from the analytic solution. The red dots denote the doublet ratios computed from the Gaussian emission profiles for each \sigmaint, which are in excellent agreement with the analytic solution.}
    \label{fig:Fig_A1}
\end{figure}

Figure~\ref{fig:Fig_A1} shows the \CIV doublet ratio \RCIV as a function of the intrinsic emission width \sigmaint. The analytic solution for \RCIV, derived from Equation~\ref{eqn_gaussian_analytic_a}, can be rewritten as
\begin{equation}
    R_{\ion{C}{IV}} = \frac{3\erf (a_{\rm int}) + \erf (b_{\rm int})}{2\erf (b_{\rm int})},
    \label{eqn_gaussian_analytic_b}
\end{equation} 
where $a_{\rm int} = 250/\sqrt{2}\sigmaint$ and $b_{\rm int} = 750/\sqrt{2}\sigmaint$.
When $3\,\sigmaint < \Delta V_{\rm K,H}$ (i.e., 500\,\kms), \RCIV has an intrinsic value of 2 since the K and H components are clearly resolved. 
In contrast, when $3\, \sigmaint > \Delta V_{\rm K,H}$, the two components become intrinsically blended, and \RCIV decreases, asymptotically approaching unity with increasing \sigmaint.
Notably, the analytic solution yields $1 \leq \RCIV \leq 2$ as a function of \sigmaint.

\section{Combined Photoionization and Resonance Scattering Model}
\label{Appendix:Photoionization}

\begin{figure*}
    \includegraphics[width=\linewidth]{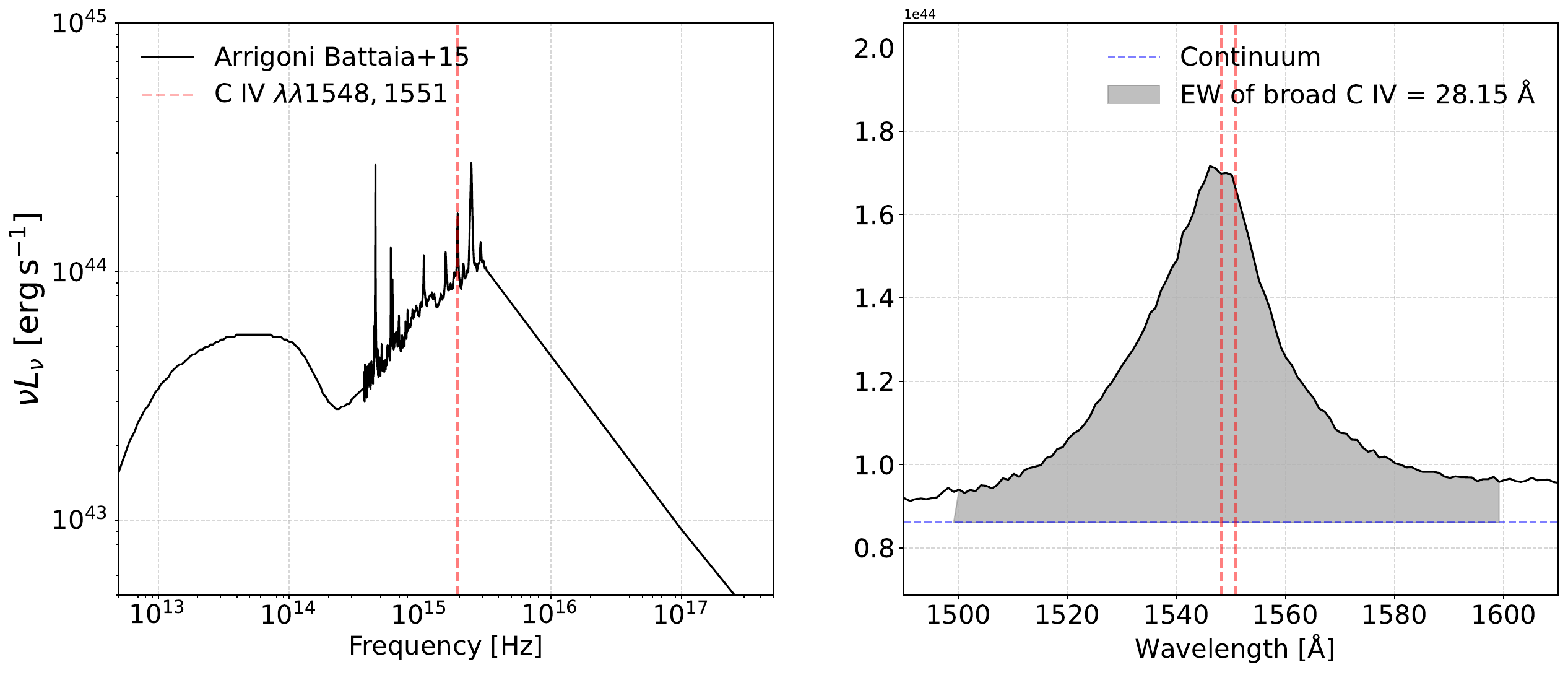}
    \caption{Spectral energy distribution (SED) of the AGN used as the incident radiation field in photoionization modeling. The solid line shows the SED from \citet{Fab2015b}, normalized to $\nu L_{\nu} = 10^{44.0}\,\rm{erg\,s^{-1}}$ at 1\,Ryd (=13.6\,eV). The left panel shows the SED in frequency space, while the right panel presents a zoom-in around the \CIV emission feature in wavelength space. The broad \CIV emission from the central source has an intrinsic width of $\sigmaint = 3000\,\kms$ and an EW of $\approx 28.2$\,\AA. The blue line marks the continuum, spanning $1500$--$1600$\,\AA, which is used for calculating the EW and modeling continuum pumping in our RT calculations. The red vertical dashed lines indicate the line centers of \CIV K- and H-lines, respectively.}
    \label{fig:Fig_B1}
\end{figure*}

In \S~\ref{sec:Spatial_distribution}, we present the spatial distributions of \CIV and \HeII emission in terms of radial surface brightness. 
Using the public photoionization code \texttt{CLOUDY}, we compute the intrinsic \CIV and \HeII surface-brightness profiles and then post-process the \CIV emission with radiative transfer simulations using \texttt{RT-scat} to quantify how resonance scattering modifies the spatial distribution. In this appendix, we describe the setup of \texttt{CLOUDY} and \texttt{RT-scat}.

\begin{figure*}
    \centering
    \includegraphics[width=\linewidth]{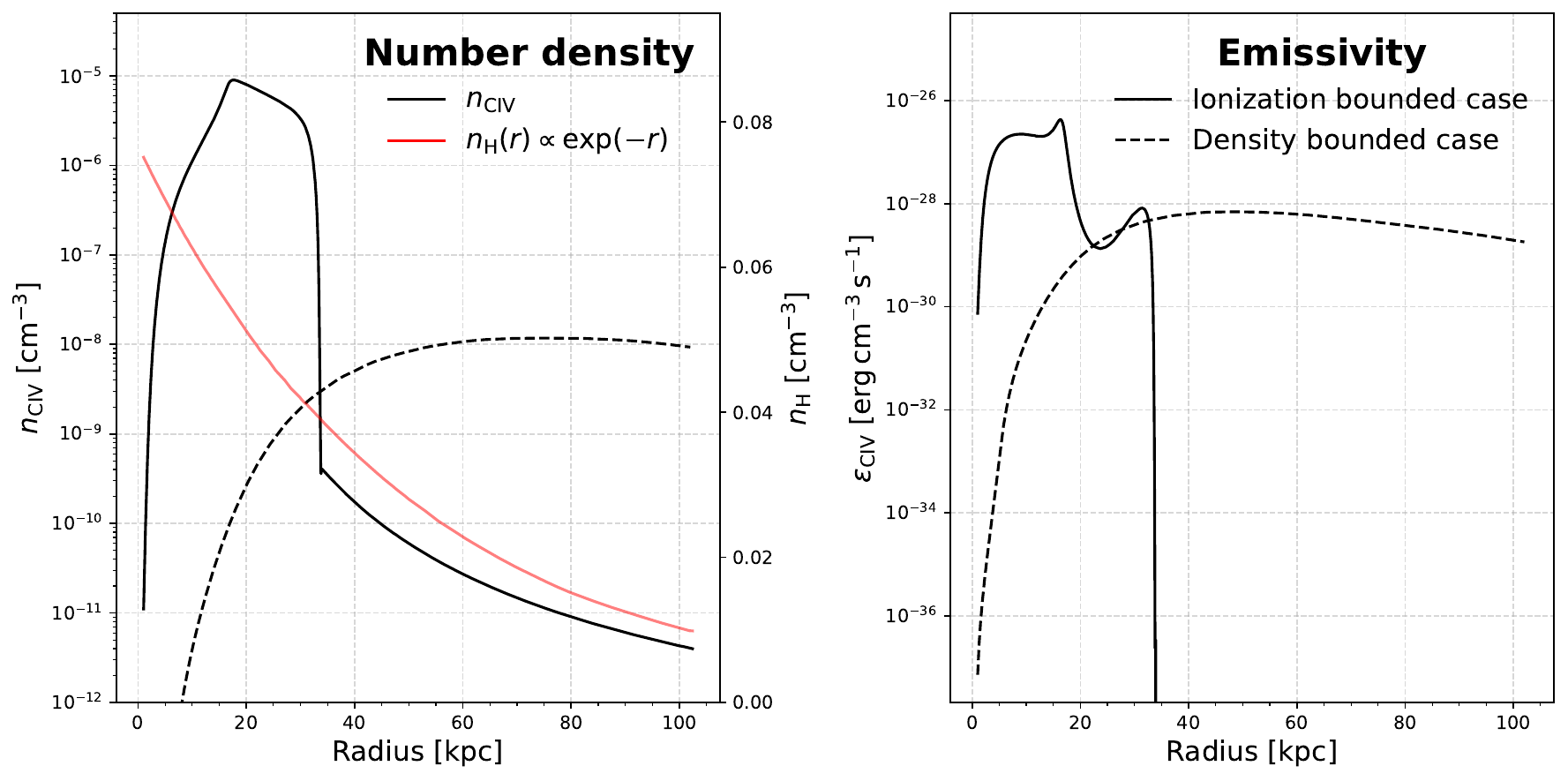}
    \caption{Radial profiles of the \CIV number density ($n_{\rm CIV}$) and emissivity ($\epsilon_{\rm CIV}$) computed with \texttt{CLOUDY}. The left panel presents the radial distributions of $n_{\rm CIV}$ (black) and $n_{\rm H}$ (red). The right panel shows $\epsilon_{\rm CIV}$ as a function of radius. In both panels, the solid and dashed lines correspond to the ionization-bounded ($\nu L_{\nu} = 10^{44.0}~{\rm erg\,s^{-1}}$) and density-bounded ($\nu L_{\nu} = 10^{46.0}~{\rm erg\,s^{-1}}$) cases, respectively.}
    \label{fig:Fig_B2}
\end{figure*}

Figure~\ref{fig:Fig_B1} shows the spectral energy distribution (SED) from \citet{Fab2015b}, which is adopted as the incident radiation field in our photoionization modeling. The SED is modeled from the stacked spectrum of \citet{Lusso2015}, corrected for absorption by the intergalactic medium. From this SED, we compute the equivalent width (EW) of $28.2$\,\AA\ for the broad \CIV emission\footnote{\citet{Lusso2015} reported a lower value of ${\rm EW} = 17.8$\,\AA\ for the \CIV emission from the same SED; our measurement is higher because we adopt a relatively wide continuum wavelength range of $1500$--$1600$\,\AA.}. This EW is measured over the wavelength range of $1500$--$1600$\,\AA, with the continuum level fitted within a window of $1450$--$1470$\,\AA\ to avoid contamination. From Gaussian fitting, we estimate an intrinsic emission width of $\sigmaint = 3000\,\kms$ for the broad \CIV emission.

Figure~\ref{fig:Fig_B2} presents the radial profiles of the \CIV number density $n_{\rm CIV}$ and emissivity $\epsilon_{\rm CIV}$, computed using \texttt{CLOUDY}. We assume that the hydrogen number density follows an exponentially decreasing radial profile,
\begin{equation}
    n_{\rm H}(r) = n_0 \exp(-r/r_e),
    \label{eqn_ap:number_density_HI}
\end{equation}
where $r_e = 50\,\rm kpc$ is the scale length, and the normalization $n_0$ is determined by
\begin{equation}
    n_0 = \frac{N_{\rm H}}{\int_{R_{\rm in}}^{R_{\rm out}} \exp(-r/r_e)\,dr}.
\end{equation}
Here, $R_{\rm in} = 1\,\rm kpc$ and $R_{\rm out} = 100\,\rm kpc$ define the inner and outer boundaries of the halo. As described in \S~\ref{sec:Spatial_distribution}, we assume a total hydrogen column density of $N_{\rm H}=10^{22}~\unitNHI$ and solar metallicity ($Z = Z_{\odot}$) throughout the halo, which allows us to reproduce the observed \HeII luminosity (i.e., $L_{\rm He\,II} = 10^{42\text{--}43}\,{\rm erg\,s^{-1}}$) within the photoionization framework. Furthermore, we consider two ionization regimes depending on whether an H~I ionization front is present (ionization-bounded) or absent (density-bounded), with the SED normalized to two distinct luminosities at 1~Rydberg ($= 13.6\,\rm eV$): $\nu L_{\nu} = 10^{44.0}$ and $10^{46.0}\,{\rm erg\,s^{-1}}$, respectively. From these assumptions, the radial distributions of $n_{\rm CIV}$ and $\epsilon_{\rm CIV}$ are derived, in contrast to the uniform density profile adopted in \S~\ref{sec:geometry}.

Figure~\ref{fig:Fig_11} presents the radial surface-brightness profiles of \CIV and \HeII emission. The intrinsic surface brightness is computed from the line emissivities shown in Figure~\ref{fig:Fig_B2} via
\begin{equation}
    {\rm SB}(r_{\rm p}) = \frac{2}{(1+z)^4} \int_{r_{\rm p}}^{R_{\rm out}}
    \frac{(\epsilon_i(r)/4\pi)\times r}{\sqrt{r^2 - r_{\rm p}^2}}\, dr,
\end{equation}
where $i$ denotes \CIV or \HeII, $r_{\rm p}$ is the projected radius spanning from 0 to 100\,kpc, $R_{\rm out} = 100\,\rm kpc$ is the outer boundary of the halo, and $z$ is the redshift of the source. Throughout this work, we adopt $z = 0$.

As discussed in \S~\ref{sec:Spatial_distribution}, the spatial distribution of the non-resonant \HeII line follows the emissivity predicted by photoionization, whereas the resonant \CIV line can be redistributed to larger radii through scattering. To incorporate resonance scattering into the photoionization framework, we post-process the \CIV emission using \texttt{RT-scat}, assuming the halo is illuminated by three distinct sources of \CIV photons.
The first source is the in-situ intrinsic emission produced within the photoionized gas, which follows the radial distribution shown in Figure~\ref{fig:Fig_B2}.
The other two sources originate from the central AGN (see Figure~\ref{fig:Fig_B1}): a broad-line region (BLR) component emitting broad \CIV photons with $\sigmaint = 3000\,\kms$, and an AGN continuum component that accounts for continuum pumping over the wavelength range of $1500$--$1600$\,\AA.
The luminosities of all three components are summarized in Table~\ref{tab:Luminosity_table}. For the \texttt{RT-scat} calculation, we adopt a velocity dispersion of $\sigmaR = 30\,\kms$ and a radial outflow velocity that varies as $v(r) = V_{\rm out}\,(r/R_{\rm H})$, where $r$ is the distance from the central source, $R_{\rm H}$ is the halo radius (see Figure~\ref{fig:Fig_3}), and the maximum outflow velocity is set to $\vout = 500\,\kms$.
The resulting surface-brightness profiles, combining contributions from all three \CIV photon sources, are presented in Figure~\ref{fig:Fig_11}.

\bsp
\label{lastpage}
\end{document}